\documentclass[prb,twocolumn,aps,superscriptaddress,showpacs]{revtex4-2}

\usepackage{amsmath,amssymb}
\usepackage{graphicx}
\usepackage{xcolor}
\usepackage{graphicx}
\DeclareGraphicsExtensions{.pdf,.eps,.png,.jpg} \graphicspath{{./figs/}}
\usepackage{dcolumn}
\usepackage{bm}
\usepackage{tabularx}
\usepackage{upgreek}
\usepackage{multirow}
\usepackage{epstopdf}
\usepackage{gensymb}

\newcommand{\rem}[1]{}
\newcommand{\red}[1]{#1}

\usepackage{amsmath}%
\usepackage{amsfonts}%
\usepackage{amssymb}%
\usepackage[utf8]{inputenc}
\usepackage{graphicx}
\usepackage{epstopdf}
\usepackage{color}
\usepackage{ulem}
\usepackage[english]{babel}
\usepackage{natbib}
\usepackage[colorlinks=true, citecolor=blue, linkcolor=blue, urlcolor=blue]{hyperref}
\usepackage{placeins}
\usepackage{booktabs}
\usepackage{enumitem}
\usepackage{subfigure}
\usepackage{multirow}
\usepackage{inputenc}
\usepackage{braket}
\usepackage{svg}
\usepackage{dcolumn}
\usepackage{bm}
\usepackage{siunitx}
\usepackage{mathtools}
\usepackage{soul}

\begin{document}

\title{Spin-orbit-entangled $J_{\rm eff}=\frac{1}{2}$ magnetism and unconventional spin freezing in the bond-disordered pyrochlore antiferromagnet NaCdCo$_2$F$_7$}

\author{A. Kancko}
\affiliation{Charles University, Faculty of Mathematics and Physics, Department of Condensed Matter Physics, Ke Karlovu 5, 121 16, Prague, Czech Republic}
\author{H. Sakai}
\affiliation{Advanced Science Research Center, Japan Atomic Energy Agency, Tokai, Ibaraki 319-1195, Japan}
\author{J. Herrero-Mart\'in}
\affiliation{ALBA Synchrotron Light Source, Carrer de la Llum 2-26, 08290 Cerdanyola del Vall\`es, Barcelona, Spain}
\author{A. Berlie}
\affiliation{ISIS Neutron and Muon Source, Rutherford Appleton Laboratory, Science and Technology Facilities Council, Chilton, Oxfordshire OX11 0QX, U.K.}
\author{M. Uhlarz}
\affiliation{Hochfeld-Magnetlabor Dresden (HLD-EMFL), Helmholtz-Zentrum Dresden-Rossendorf (HZDR), 01328
Dresden, Germany}
\author{T. Haidamak}
\affiliation{Charles University, Faculty of Mathematics and Physics, Department of Condensed Matter Physics, Ke Karlovu 5, 121 16, Prague, Czech Republic}
\author{M. Klicpera}
\affiliation{Charles University, Faculty of Mathematics and Physics, Department of Condensed Matter Physics, Ke Karlovu 5, 121 16, Prague, Czech Republic}
\author{Y. Tokunaga}
\affiliation{Advanced Science Research Center, Japan Atomic Energy Agency, Tokai, Ibaraki 319-1195, Japan}
\author{R. H. Colman}
\affiliation{Charles University, Faculty of Mathematics and Physics, Department of Condensed Matter Physics, Ke Karlovu 5, 121 16, Prague, Czech Republic}

\date{\today}

\begin{abstract}
Bond disorder in frustrated pyrochlore antiferromagnets can give rise to fundamentally different quantum ground states depending on the nature of the local magnetic moments. Here, we show that the bond-disordered \red{$J_{\rm eff}=\frac{1}{2}$} pyrochlore \red{antiferromagnet} NaCdCo$_2$F$_7$ realizes an unconventional spin-glass-like state with continued dynamics, in stark contrast to its isostructural $S=\frac{1}{2}$ NaCdCu$_2$F$_7$ counterpart. High-field magnetization and Co $L_{2,3}$-edge XAS/XMCD establish spin-orbit-entangled $J_{\rm eff}=\frac{1}{2}$ Co$^{2+}$ moments with a substantial unquenched orbital contribution, consistent with local $XY$ anisotropy seen in the isostructural Na$A''$Co$_2$F$_7$ ($A''$ = Ca, Sr) analogues. $\mu$SR and $^{23}$Na NMR measurements reveal progressive slowing of spin fluctuations below $\sim10$ K, culminating in a partially frozen state with persistent low-temperature dynamics that deviates from a canonical spin glass. Comparison with the isostructural bond-disordered pyrochlore NaCdCu$_2$F$_7$, which realizes a random-singlet state, reveals a fundamentally different response of spin-orbit-entangled Co$^{2+}$ moments to bond disorder. These results identify spin-orbit coupling as a key ingredient governing the fate of bond-disordered frustrated pyrochlore magnets.

\end{abstract}

\pacs{}

\maketitle

\section{Introduction}

Geometrically frustrated magnets provide a fertile platform for the study of unconventional magnetic ground states, including spin glasses, spin ices, and spin liquids \cite{Ramirez1994,Greedan2001,Balents2010,Ramirez2025,Klicpera2026}. Among the most prominent realizations are pyrochlore magnets, in which \rem{antiferomagnetically-coupled} \red{antiferromagnetically coupled} magnetic ions occupy a three-dimensional network of corner-sharing tetrahedra that gives rise to strong geometric frustration, hindering the development of trivial magnetic order. In these systems, subtle differences in electronic configuration, spin-orbit coupling, local crystal-field environment, exchange anisotropy and disorder can stabilize fundamentally different ground states \cite{Greedan2006,Reig-I-Plessis2021}.

The rare-earth family of $A^{3+}_2B^{4+}_2$O$_7$ pyrochlore oxides has been extensively studied due to the availability of large single crystals in many cases, enabling detailed microscopic investigations of their rich variety of exotic magnetic ground states \cite{Gardner2010,Klicpera2026, Stasko2024b,Stasko2024a}. In these materials, spin-orbit coupling and crystal-field effects play a central role in forming strong Ising or XY anisotropy, stabilizing diverse magnetic phases such as spin-ice or unconventional ordered states \cite{Snyder2004,Ross2011,Champion2003,Zhitomirsky2012,Stasko2024c}. 
However, the weak exchange energy scale of localized $4f$ rare-earth moments typically confines the relevant magnetic phenomena to extremely low temperatures, making experimental access to their low-energy physics challenging.

Transition-metal fluoride pyrochlores $A'^+A''^{2+}B^{2+}_2$F$_7$ \cite{Hansler1970,Reig-I-Plessis2021,Krizan2014,Krizan2015,Krizan2015Ni,Sanders2017,Kancko2023,Kancko2025,Kancko2026}, conversely, host divalent magnetic $3d$ transition-metal ions on the pyrochlore $B$ site, leading to substantially stronger superexchange interactions mediated by fluoride ligands. Charge balancing and structure stability constraints, however, \rem{result in chemical disorder on the nonmagnetic $A$-site sublattice through mixing of monovalent $A'^+$ and divalent $A''^{2+}$ cations~\cite{Song2020}} \red{require
random mixing of monovalent $A'^{+}$ and divalent $A''^{2+}$ cations on
the nonmagnetic pyrochlore $A$ sublattice \cite{Song2020}}. \rem{This cationic disorder, in combination with the $A'^+/A''^{2+}$ ionic radii mismatch, introduces randomness in the magnetic superexchange pathways coupling neighboring $B$-site ions.} \red{The associated $A'^+/A''^{2+}$ ionic-size mismatch then introduces local
structural distortions, which perturb the $B$--F--$B$ superexchange
pathways and thereby generate a spatial distribution of magnetic exchange
interactions. We refer to this exchange randomness as magnetic bond
disorder.} This weak magnetic bond disorder on the pyrochlore lattice is expected to precipitate spin-glass freezing at a temperature that is commensurate with the exchange variation scale $k_{\rm B}T_f = \sqrt{8/3} \Delta$ \cite{Saunders2007,Andreanov2010}.

Experimental studies of fluoride pyrochlores broadly support this disorder-driven freezing scenario. A vast majority of the Na$A''$$B_2$F$_7$ family ($A''$ = Ca, Sr, Cd; $B$ = Co, Ni, Mn, Fe) \cite{Krizan2014,Krizan2015,Krizan2015Ni,Sanders2017, Kancko2023,Kancko2025} undergo spin-glass-like freezing at temperatures $T_f = 2-4$ K, substantially lower than their mean-field exchange energy scales ($|\theta_{\rm CW}| \sim 70-140$ K). \red{Within the isostructural Na$A''$Co$_2$F$_7$
($A''$ = Ca, Sr, Cd) series, the role of this disorder is further
supported by a systematic correlation between the $A$-site ionic-size
mismatch and the spin-freezing temperature: $T_f$ increases from 2.4 K for $A''$ = Ca to 3.0 K for $A''$ = Sr and 4.0 K for $A''$ = Cd, while
the corresponding ionic-size mismatch increases from approximately
5.2\% to 6.6\% and 7.0\%, respectively \cite{Kancko2023}.}

Despite the apparent spin freezing, local probe measurements on the $S=1$ NaCaNi$_2$F$_7$ revealed persistent low-temperature spin dynamics well below its freezing temperature \cite{Cai2018}, while inelastic neutron scattering uncovered spin-liquid-like correlations, including a continuum of magnetic scattering with low-energy pinch points \cite{Plumb2019, Zhang2019}. These observations suggest that geometric frustration inhibits complete static freezing and gives rise to unconventional glassy states rather than canonical spin glasses \cite{Zhou2008,Silverstein2014}. \red{Here, we use the term ``unconventional spin glass'' to refer
to a partially frozen, dynamically inhomogeneous state in which
quasi-static local magnetism coexists with persistent spin fluctuations
well below the bulk freezing temperature, in contrast to the predominantly
static frozen state of a canonical spin glass.} Strikingly, the recently-synthesized $S=\frac{1}{2}$ \red{Heisenberg pyrochlore antiferromagnet} NaCdCu$_2$F$_7$ stands out as an important exception that realizes a disorder-driven random-singlet state rather than an unconventional spin glass \cite{Kancko2026}. It is  characterized by power-law scaling of bulk and local-probe thermodynamic quantities, universal scaling collapse, and the absence of global spin freezing. An important question arises about what governs the fundamentally different response of fluoride pyrochlores to comparable bond disorder.

Among fluoride pyrochlores, Co$^{2+}$ systems are particularly intriguing due to the strong spin-orbit coupling of the high-spin $3d^7$ electronic configuration. Single-ion inelastic neutron scattering (INS) studies of NaCaCo$_2$F$_7$ and NaSrCo$_2$F$_7$ established a well-isolated $J_{\rm eff}=\frac{1}{2}$ Kramers doublet characterized by a strongly anisotropic local XY-type $g$ tensor and anisotropic exchange interactions \cite{Ross2017}. Furthermore, diffuse neutron scattering and magnetic pair distribution function studies revealed the emergence of short-range ordered $XY$ clusters at low temperatures, indicative of highly anisotropic short-range magnetic correlations \cite{Ross2016,Frandsen2017}. Complementary $^{19}$F and $^{23}$Na NMR measurements further demonstrated persistent low-energy spin fluctuations and unconventional freezing dynamics in NaCaCo$_2$F$_7$ \cite{Sarkar2017}. Together, these observations suggest that spin-orbit-entangled Co$^{2+}$ moments may respond to bond disorder in a fundamentally different manner from nearly isotropic Heisenberg systems. However, analogous microscopic information remains unavailable for NaCdCo$_2$F$_7$, where the strong neutron absorption of natural Cd precludes detailed neutron-scattering investigations of both the crystal-field level scheme and the nature of short-range magnetic correlations.

In this work, we combine high-field magnetization, Co $L_{2,3}$-edge X-ray absorption and magnetic circular dichroism (XAS/XMCD), muon spin relaxation ($\mu$SR), and $^{23}$Na nuclear magnetic resonance (NMR) measurements to investigate the magnetic ground state of NaCdCo$_2$F$_7$. High-field magnetization and XAS/XMCD establish spin-orbit-entangled Co$^{2+}$ moments with a substantial unquenched orbital contribution, consistent with the local $XY$ anisotropy established in the isostructural Na$A''$Co$_2$F$_7$ ($A''$ = Ca, Sr) analogues. Furthermore, $\mu$SR and $^{23}$Na NMR reveal a progressive slowing down of spin fluctuations into an unconventional partially frozen state with persistent low-temperature dynamics. Comparison with the random-singlet $S=\frac{1}{2}$ Heisenberg pyrochlore antiferromagnet NaCdCu$_2$F$_7$ reveals a fundamentally different response to bond disorder, identifying spin-orbit coupling and magnetic anisotropy as key ingredients governing the fate of bond-disordered pyrochlore antiferromagnets.

\section{Experimental methods}

Single crystals of NaCdCo$_2$F$_7$ were grown in an optical floating zone furnace by pre-melting a stoichiometric mix of precursor binary fluorides in a tubular graphite crucible, then hanging the pre-reacted polycrystalline rod for a standard floating-zone growth in a high-pressure dynamic argon atmosphere. Details of the synthesis process are covered in Ref. \cite{Kancko2023}. Crystals were then oriented for directional property measurements using an in-house Laue diffractometer.

High-field isothermal magnetization was measured on oriented single crystals of NaCdCo$_2$F$_7$ in the High Magnetic Field Laboratory (Hochfeld-Magnetlabor Dresden, HLD) in the Helmholtz-Zentrum Dresden-Rossendorf (HZDR) in Dresden, Germany, using a pulsed 60 T magnet with a cryostat setup allowing the temperature range $T = 1.37 - 100$ K \cite{Skourski2011}. The magnetic field was applied along the cubic high-symmetry directions $[100], [110]$ and $[111]$. Calibration of pulsed field data was carried out by normalization against in-house MPMS and PPMS datasets.

XAS/XMCD spectra were measured on an unoriented single crystal of NaCdCo$_2$F$_7$ at the BL29 BOREAS soft X-ray beamline at the ALBA synchrotron in  Cerdanyola del Vall\`es, Barcelona, Spain \cite{Barla:vv5144}. The circularly polarized XAS spectra of the Co$^{2+}$ $L_{2,3}$ edge were collected between $765-810$ eV at 300 K in the normal incidence geometry, using the total electron yield (TEY) mode. A magnetic field of 6 T was applied along the beam direction, using the vector magnet HECTOR. A statistical average of two octets (C+C-C-C+C-C+C+C-) was collected and used for the analysis. The XMCD signal was then calculated as the difference of the XAS spectra with negative and positive helicity. 

ZF and LF-$\mu$SR measurements were performed at the ISIS Neutron and Muon Source at the Rutherford Appleton Laboratory in Harwell Oxford, United Kingdom. \cite{Colman2024} The MUSR instrument was used, utilising a helium cryostat setup for temperatures between $T = 1.7-50$ K, allowing applied longitudinal fields between $\mu_0H_{\rm LF} = 0-3000$ G. The obtained spectra were analyzed using WiMDA \cite{Pratt2000} and Mantid \cite{Arnold2014}.

NMR measurements were performed using a phase-coherent pulsed spectrometer with a superconducting NMR magnet. A $^4$He variable temperature insert (VTI) was utilized between $T = 1.7-60$ K. A custom-wound  excitation coil from silver wire was used.
The single crystal of NaCdCo$_2$F$_7$ was inserted into the NMR coil, and the coil was mounted on a sample stage equipped with a two-axis rotation mechanism.
A small piece of pure Al foil was mounted next to the single-crystal sample as a field marker, and the applied magnetic field was calibrated using its $^{27}$Al NMR signal.
Frequency-swept $^{23}$Na spectra were collected in a constant field $\mu_0 H = 2.21331$ T, with the radio-frequency (r.f.) circuit tuned and matched at each point. Nuclear spin echoes were generated using a standard 90$\degree$--180$\degree$ pulse sequence, with a first-pulse duration of 2 - 3 $\mu$s, where the r.f. power for nuclear spin excitation was optimized at each NMR spectral peak. The separation $\tau$ between the first and second pulse was typically 20 - 30 $\mu$s. Echoes were accumulated and Fourier transformed to obtain the spectra.

\section{Results}

\subsection{High-field magnetization}

\begin{figure}[t]
\begin{center}
\includegraphics[scale=1]{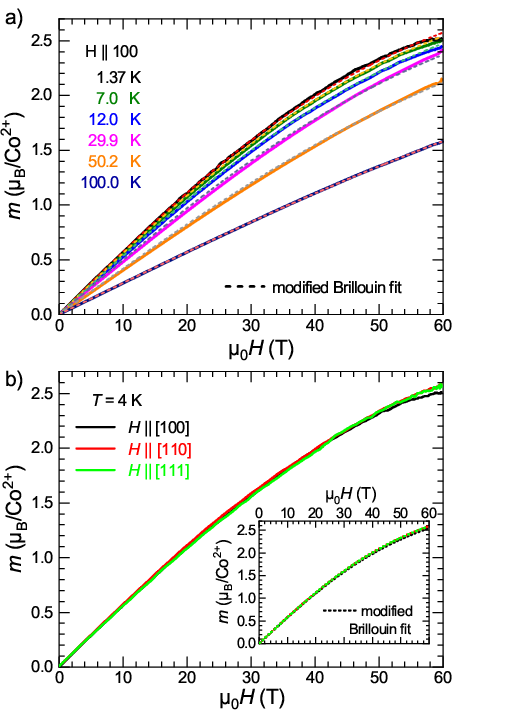}
\caption{(a) Temperature dependence of pulsed-field magnetization of an oriented crystal of NaCdCo$_2$F$_7$, with $H \parallel [100]$. Fits to Eq. (\ref{eq1}) are shown as dashed lines. (b) Directional dependence of magnetization at $T=4$ K, with field applied along [100], [110] and [111] directions. Fits to Eq. (\ref{eq1}) are shown separately in the inset.} 
\label{fig1} 
\end{center}
\end{figure}

High-field isothermal magnetization $M(H)$ measured in pulsed fields up to 60 T, applied along the [100] direction at temperatures between $1.37 - 100$ K, is shown in Fig. \ref{fig1}(a). No magnetization plateaus or field-induced transitions are observed, in contrast to recent theoretical predictions for ideal disorder-free $S=\frac{1}{2}$ pyrochlore Heisenberg antiferromagnets under magnetic field \cite{Pal2019,Schafer2020,Hagymasi2022}. Instead, the magnetization increases smoothly with field and exhibits a slow Brillouin-like polarization that does not reach full saturation even at the largest field-to-temperature ratio of $\mu_0H/T \sim 44$ \red{T\,K$^{-1}$} achieved at 60 T and 1.37 K. Previous low-field  SQUID magnetization measurements of NaCdCo$_2$F$_7$ \cite{Kancko2023} revealed strong antiferromagnetic  exchange interactions characterized by a Curie-Weiss temperature of $\theta_{\rm CW} = -108(1)$ K. Using the mean-field relation $\theta_{\rm CW} = 2zJ(J+1)\mathcal{J}/3k_B$, where $z=6$ is the nearest-neighbor coordination number of the pyrochlore lattice and $J=J_{\rm eff} = \frac{1}{2}$, we estimate an average nearest-neighbor exchange coupling of $\mathcal{J} \approx -36$ K ($\approx -3.1$ meV). The strong antiferromagnetic exchange interactions combined with the geometrical frustration of the pyrochlore lattice hinder full magnetic saturation even by 60 T, which in a non-interacting Heisenberg system of spins would be expected by several Tesla \cite{Zeisner2019}. 

Single-ion inelastic neutron scattering (INS) studies of the isostructural Na$A''$Co$_2$F$_7$ ($A''$ = Ca, Sr) family revealed a well-isolated spin-orbit-coupled $J_{\rm eff} = \frac{1}{2}$ Kramers doublet ground state with a strongly $XY$-like $g$ tensor ($g_{\rm xy} = 6.08$ and $g_z = 1.87$), separated from the first excited doublet by 28.05(2) and 29.31(2) meV (or 325.5(2) and 340.1(2) K) in $A''$ = Ca and Sr, respectively \cite{Ross2017}. While neutron scattering studies are not possible in NaCdCo$_2$F$_7$ due to the very large neutron absorption cross-section of natural Cd, the two-level Schottky anomaly fit of magnetic specific heat revealed a possible first excited state at around 33.9(5) meV (or 393(6) K) \cite{Kancko2023}. Consistently, the magnetic entropy approaches Rln(2) by 70 K in all $A''$ = Ca, Sr and Cd cobalt pyrochlores \cite{Krizan2014,Krizan2015,Kancko2023}, confirming that only the ground-state $J_{\rm eff} = \frac{1}{2}$ Kramers doublet is thermally populated in the temperature range of the present measurements. It is therefore possible to model the low-temperature $M(H)$ data using a modified $J_{\rm eff} = \frac{1}{2}$ Brillouin function with an effective temperature $T_{\rm eff} = T-T_0$ accounting for the exchange interactions between Co$^{2+}$ spins \cite{Heiman1984,Cisowski1999}:

\begin{equation}
    M_{\mathbf d}(H,T) = M_{\mathbf{d}}^{\rm sat} \mathrm{tanh} \left( \frac{g_{\mathbf{d}} \mathrm{\mu_B \mu_0} H}{2\mathrm{k_B}(T-T_0)} \right) 
    \label{eq1}
\end{equation}

where $M_{\mathbf{d}}^{\rm sat} = g_{\mathbf{d}}\mu_{\rm B}/2$ is the saturation moment along the field direction $\mathbf{d}= \vec{\mathbf{H}}/|\vec{\mathbf{H}}|$, $g_{\mathbf{d}}$ is the projection of the anisotropic single-ion $g$-tensor onto the field direction averaged over the four sites on the tetrahedron, and $T_0$ is a phenomenological mean-field parameter that reflects the strength of antiferromagnetic exchange interactions. The fits are shown in Fig. \ref{fig1}(a) as dashed lines, and the extracted parameters are summarized in Table \ref{tab1}. 

A global fit of $M(H)$ was performed between $1.37 - 100$ K along $H \parallel [100]$, with the effective $g$-factor shared for all temperatures and $T_0$ fitted independently. The fit yields $g_{[100]} = 6.563(1)$, corresponding to a saturation moment of $M_{[100]}^{\rm sat} = 3.28$ $\mu_{\rm B}$/Co$^{2+}$. The effective moment evaluated as $\mu_{\rm eff} = g_{[100]} \sqrt{J_{\rm eff}(J_{\rm eff}+1)} \mu_{\rm B} = 5.68$ $\mu_{\rm B}$/Co$^{2+}$ is in good agreement with the orientationally-averaged paramagnetic value of 5.4(1) $\mu_{\rm B}$/Co$^{2+}$  extracted from the Curie-Weiss analysis of inverse susceptibility \cite{Kancko2023} between $100-350$ K. The phenomenological mean-field interaction parameter $T_0$ is nearly temperature independent for temperatures between $1.37-50.2$ K, with  $T_0^{[100]} \approx -121(3)$ K, consistent with the scale of mean-field interactions given by the Curie-Weiss temperature $\theta_{\rm CW} = -108(1)$ K. The large deviation in the 100 K value, $T_0^{[100]}$(100 K) $=-151.21(6)$ K, may be attributed to a non-negligible thermal population of the first excited Kramers doublet. 

Using the fitted value $g_{[100]} = 6.563(1)$, we can estimate the saturation field from the competition between antiferromagnetic exchange and the Zeeman energy $\mu_0 H_{\mathrm sat} = \frac{z |\mathcal{J}|}{2 g \mu_{\rm B} J_{\rm eff}}$ \cite{Landee2014}. For $J_{\rm eff} = \frac{1}{2}$, $z = 6$, and $|\mathcal{J}| \approx 36~\mathrm{K}$, this gives $\mu_0 H_{\rm sat} \sim 49 ~\mathrm{T}$. Instead, only $\sim \!76$ \% of the expected saturated moment is observed by 60 T at the lowest temperature, as geometrical frustration, quantum fluctuations, and anisotropic exchange interactions all tend to increase the effective saturation field relative to this simple mean-field estimate.

The directional dependence of $M(H)$ at 4 K is shown in Fig. \ref{fig1}(b) for fields applied along the cubic high-symmetry directions $[100]$, $[110]$, and $[111]$. For clarity, fits to Equation \ref{eq1} are shown separately in the inset of Fig. \ref{fig1}(b) due to the large overlap with the experimental data. The magnetization curves appear strongly isotropic, suggesting Heisenberg-like magnetism with only weak signs of anisotropy observed by 60 T, which could come from instrumental effects near the pulsed magnet limit. The extracted parameters for $H \parallel [110]$ and $[111]$ are essentially identical within the error, with $g_{[110]} = g_{[111]} = 6.566(1)$ and $T_0^{[110]} \sim T_0^{[111]} \sim -120$ K. In contrast, an independent fit of the 4 K data along the [100] direction yields slightly smaller values of $g_{[100]} = 6.423(1)$ and $T_0^{[100]} = -116.21(6)$ K, as compared to the values extracted from the global fit of the $M_{[100]}(H)$ dataset between $1.37 - 100$ K. The difference arises largely from the sharing of the $g$ factor across the whole temperature range, inflating its value while simultaneously influencing the fitted $T_0$. 

Our results are consistent with the seemingly isotropic high-field magnetization curves of NaCaCo$_2$F$_7$ \cite{Zeisner2019}, despite its strongly anisotropic XY-type single-ion $g$ tensor ($g_{xy}/g_z \sim 3)$, where one would naively expect $|M_{[100]}| < |M_{[111]}| < |M_{[110]}|$. The apparent Heisenberg behavior arises because the local XY anisotropy of each of the four $J_{\rm eff} = \frac{1}{2}$ Co$^{2+}$ spins on a tetrahedron (each with a local $\langle 111\rangle$ anisotropy axis) averages out when projected onto the global field direction, leading to isotropic magnetization curves even in the presence of strong local single-ion anisotropy. The effective projected $g$-factor can be calculated from an XY-type anisotropic $g$-tensor as $g_{\rm eff} = \sqrt{\frac{1}{3}(2g_{xy}^2 + g_z^2 ) }$, yielding $g_{\rm eff} = 5.08$ for NaCaCo$_2$F$_7$. Our results suggest that NaCdCo$_2$F$_7$ with $g_{\rm eff} \sim 6.56$ most likely exhibits a greater local XY-type anisotropy in the form of a larger dominant $g_{xy}$ component.
\cite{Hoffmann2013}

\begin{table}[h]
\caption{Parameters obtained from $M(H)$ fits to Equation \ref{eq1}.}
\renewcommand{\arraystretch}{1.5}
\setlength{\tabcolsep}{3pt}
\begin{tabular}{c|c|c|c}
Orientation                          & $T$ (K) & $g_{\bf d}$ & $T_0$ (K)  \\ \hline
\multirow{7}{*}{$H \parallel [100]$} & 1.37    & \multirow{7}{*}{6.563(1)}       & -123.69(3) \\
 & 4.0     &                                 & -122.68(3) \\
  & 7.0     &                                 & -121.13(3) \\
  & 12.0    &                                 & -122.41(4) \\
 & 29.9    &                                 & -114.66(4) \\
  & 50.2    &                                 & -121.84(4) \\
 & 100.0   &                                 & -151.35(6) \\
 $H \parallel [100]$                  & 4.0     & 6.423(1)                        & -116.21(6) \\
$H \parallel [110]$                  & 4.2     & 6.566(1)                        & -119.89(6) \\
$H \parallel [111]$                  & 4.2     & 6.566(1)                        & -120.12(6)
\end{tabular}
\label{tab1}
\end{table}

\subsection{X-ray magnetic circular dichroism}

To probe the microscopic origin of the large effective magnetic moment observed in NaCdCo$_2$F$_7$ and related cobalt-based fluoride pyrochlores \cite{Kancko2023,Krizan2014,Krizan2015}, \rem{as well as the role of spin-orbit coupling (SOC) in their local magnetic anisotropy \cite{Ross2017,Ross2016},} \red{and in particular
the role of spin-orbit coupling (SOC) in the Co$^{2+}$ magnetic degrees
of freedom and its connection to the local magnetic anisotropy
established in the isostructural Na$A''$Co$_2$F$_7$
($A''=$ Ca, Sr) analogues \cite{Ross2016,Ross2017},} X-ray magnetic circular dichroism (XMCD) measurements were performed at the Co$^{2+}$ $L_{2,3}$ absorption edge. 

Single-crystal X-ray diffraction studies have shown \cite{Kancko2023} that Co$^{2+}$ fully occupies the pyrochlore $16c$ Wyckoff site with a local $D_{3d}$ symmetry, coordinated by six fluorine ions forming slightly distorted CoF$_6$ octahedra compressed along the local $\langle 111 \rangle$-type ($C_3$) axis. This small trigonal compression produces a deviation of the F-Co-F bond angles from the ideal octahedral value of 90$\degree$ by $\pm \delta_{\rm F-Co-F} = 8.12(5) \degree$. Such a local symmetry lowering is expected to partially lift the cubic $t_{2g}$ degeneracy as $\ket{t_{2g}} \rightarrow \ket{a_{1g}} \bigoplus \ket{e_g^\pi}$, while leaving the $\ket{e_g}$ orbitals untouched \cite{DiScala2026}. Together with the strong spin–orbit coupling of high-spin Co$^{2+}$, this leads to the stabilization of six Kramers doublets, with a $J_{\rm eff}=\frac{1}{2}$ doublet as the ground state \cite{Abragam1951,Holden1971,Liu2018,Piwowarska2019}\rem{, resulting in strongly locally-anisotropic magnetic properties.}. \red{Such spin--orbit-entangled
Co$^{2+}$ moments can exhibit strongly anisotropic local magnetic
properties.}

\begin{figure}[htbp]
    \centering
    \includegraphics[width=1\linewidth]{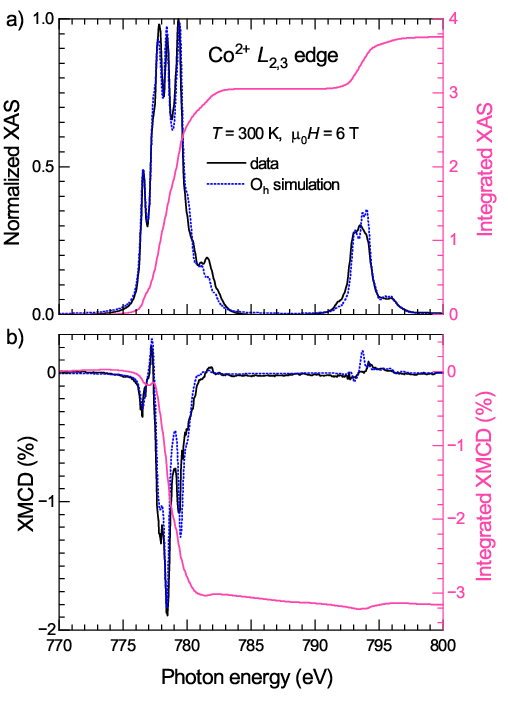}
    \caption{(a) Co$^{2+}$ $L_{2,3}$ XAS and (b) XMCD \rem{data} \red{spectra} measured at $T=300$ K in 6 T field, both normalized to the maximum $L_3$ intensity, shown together with the $O_h$ \rem{and $C_{3v}$ crystal-field simulations} \red{crystal-field multiplet calculations} from Crispy. The \red{XAS/XMCD} integration curves used for the sum-rules analysis \rem{of XAS/XMCD} are shown on the right axes.}
    \label{figXMCDsimul}
\end{figure}

The X-ray absorption spectrum (XAS), measured in the TEY mode at $T= 300$ K in an applied field of $\mu_0H= 6$ T, and normalized to the \rem{highest} \red{maximum} intensity of the $L_3$ peak, can be seen in Fig. \ref{figXMCDsimul}(a). The spectrum was corrected by a standard linear background and edge step function subtraction, and is defined as XAS = $\mu^-(E)+\mu^+(E)$, where $\mu^+(E)$ and $\mu^-(E)$ correspond to the absorption spectra measured using left- and right-circularly polarized X-rays, respectively. 

The XAS exhibits the characteristic multiplet structure of octahedrally coordinated high-spin Co$^{2+}$ \red{($3d^7$, $S=3/2$)}, with intense $L_3$ and $L_2$ lines centered around 778 eV and 793 eV, respectively. The XMCD spectrum, defined as XMCD = $\mu^-(E)-\mu^+(E)$, is shown in Fig. \ref{figXMCDsimul}(b) after normalization to the maximum intensity of the $L_3$ line. A pronounced negative dichroic signal is observed at the $L_3$ edge, while the corresponding $L_2$ contribution is considerably weaker, indicating a sizable unquenched orbital contribution to the Co$^{2+}$ magnetic moment.

To assess the influence of the small local trigonal distortion on the Co$^{2+}$ electronic structure, \rem{ligand-field} \red{crystal-field} multiplet calculations \cite{Haverkort2012,deGroot1990} of the Co $L_{2,3}$ XAS/XMCD spectra were performed using the Crispy \cite{retegan_crispy} interface to Quanty \cite{quanty}. Two crystal-field models were considered: (i) an ideal octahedral $O_h$ environment, and (ii) a weakly distorted trigonal environment described within the $C_{3v}$ point symmetry.  Although the true local symmetry of the Co$^{2+}$ site is $D_{3d}$, such a symmetry option is not implemented in Crispy for Co $L_{2,3}$ calculations. The available $C_{3v}$ crystal field nevertheless captures the essential trigonal symmetry lowering from cubic symmetry and therefore provides a reasonable approximation for modeling the qualitative effects of trigonal distortion on the XAS/XMCD line shapes.

The dominant octahedral crystal field was parametrized by the cubic splitting parameter $10Dq = 0.9$ eV, which controls the energy separation between the lower $t_{2g}$ and upper $e_g$ manifolds. Electron-electron interactions were described using uniformly reduced Slater-Condon integrals scaled by factors $F^k = 0.75$ and $G^k = 0.78$ relative to their free-ion Hartree-Fock values. Such reductions are commonly employed in transition-metal multiplet calculations \cite{Haverkort2012,deGroot1990} to phenomenologically account for intra-atomic screening and covalency effects, which decrease the effective Coulomb and exchange interactions relative to the ionic limit. Spin–orbit coupling was retained at its atomic value (scale factor $\zeta=1$). To model the experimentally established trigonal distortion of CoF$_6$ octahedra \cite{Kancko2023}, additional trigonal crystal-field parameters $D_\sigma$ and $D_\tau$ were introduced within the $C_{3v}$ crystal-field description  \cite{Hempel1976, Varga1972}, which describe the splitting of the cubic $\ket{t_{2g}}$ manifold into $\ket{a_{1g}}$ and $\ket{e_g^\pi}$ states under distortion along the local threefold axis. \rem{In the present calculations, a small finite trigonal field of $D_\sigma = 0.06$ eV and $D_\tau=0$ was sufficient to reproduce the main experimental spectral features. The Hamiltonian parameters used in the $O_h$ and $C_{3v}$ simulations are summarized in Tab. \ref{tabXMCD}.} \red{The influence of
the trigonal distortion was investigated by systematically varying
$D_\sigma$ and $D_\tau$, as described in the SM
\cite{Supplementary}. Since these calculations are intended to assess
the sensitivity of the calculated spectra to weak trigonal
crystal-field perturbations rather than to determine a unique set of crystal-field parameters, the $O_h$ calculation is retained as the
minimal description in the main text. The Hamiltonian parameters used
for this $O_h$ calculation are summarized in
Tab.~\ref{tabXMCD}.} 

As shown in Fig.~\ref{figXMCDsimul}(a,b),
\rem{both the $O_h$ and weakly distorted $C_{3v}$ calculations reproduce}
\red{the $O_h$ calculation reproduces}
the overall multiplet structure and the shape of the XMCD line reasonably
well, including the dominant negative dichroic response at the $L_3$ edge
characteristic of Co$^{2+}$ with a substantial orbital contribution.
\red{Systematic variations of $D_\sigma$ and $D_\tau$ presented in the
SM \cite{Supplementary} show that the XMCD response is more sensitive to
the trigonal crystal field than the XAS line shape. Comparison using
normalized residual factors favors a crystal field close to the $O_h$
limit, although weak trigonal perturbations remain compatible with the
experimental spectra.}
\rem{The similarity of the two simulations indicates that the trigonal
distortion acts primarily as a weak perturbation to the dominant octahedral
crystal field, consistent with the relatively small structural deviation
of the CoF$_6$ octahedra from ideal cubic symmetry reported by
single-crystal diffraction studies.}
\red{This is consistent with the relatively small structural deviation of
the CoF$_6$ octahedra from ideal cubic symmetry reported by single-crystal
diffraction studies.}
\rem{However, given the limited signal-to-noise ratio of the experimental
XMCD data and the absence of a systematic fitting procedure, the present
calculations should be regarded as qualitative rather than a unique
quantitative determination of the crystal-field parameters. Nevertheless,
the multiplet calculations demonstrate that}
\red{The multiplet calculations therefore support the conclusion that}
the Co$^{2+}$ electronic structure in NaCdCo$_2$F$_7$ is well described
by a predominantly octahedral crystal field,
\rem{with a small trigonal perturbation} consistent with the expected
local $J_{\rm eff}=\frac{1}{2}$ description arising from the interplay
of crystal-field splitting and spin-orbit coupling.

Quantitative estimates of the orbital magnetic moment $m_L = \langle L_z \rangle \mu_{\rm B}$ and the effective spin magnetic moment $m_{S_{\rm eff}} = 2\langle S_{\rm eff} \rangle \mu_{\rm B}$ were obtained using the sum rules analysis \cite{Thole1992,Carra1993,Chen1995}. For the Co $L_{2,3}$ edge, the expectation value of the orbital angular momentum's projection $\langle L_z \rangle$ is given by

\begin{align}
\langle L_z \rangle &= -\frac{4}{3} \frac{\int_{L_3+L_2} \mathrm{XMCD} ~dE}{\int_{L_3+L_2}\mathrm{XAS}~ dE} (10-\langle n_{3d}\rangle) \nonumber \\ 
   &= -\frac{4q(10-\langle n_{3d}\rangle)}{3r},
\end{align}

where $q$ is the integrated XMCD signal over the $L_3+L_2$ edges, $r$ is the integrated XAS intensity and $\langle n_{3d} \rangle$ is the expectation value of the number of $3d$ electrons. The effective spin expectation value, $ \langle S_{\rm eff} \rangle = \langle S_z\rangle + \frac{7}{2}\langle T_z\rangle$, which includes the magnetic dipole operator contribution $\langle T_z\rangle$, is evaluated as

\begin{align}
\langle S_{\rm eff} \rangle &= -\frac{3\int_{L_3} \mathrm{XMCD} ~dE - 2\int_{L_2} \mathrm{XMCD} ~dE} {\int_{L_3+L_2} \mathrm{XAS} ~dE} (10-\langle n_{3d}\rangle) \nonumber \\ 
   & = -\frac{(3p-2q)(10-\langle n_{3d}\rangle)}{r},
\end{align}

where $p$ denotes the integrated XMCD intensity over the $L_3$ edge. The magnetic dipole operator $\langle T_z\rangle$ measures the anisotropy of the spin density distribution -- it vanishes for spherical charge distributions, and should remain small but non-zero in a weakly trigonally distorted environment. 

Assuming an ionic, high-spin Co$^{2+}$ $3d^7$ electronic configuration
without any mixed-valence effects, corresponding to
$\langle n_{3d}\rangle=7$, and using the $p$, $q$, and $r$ values
extracted from the integration curves shown in
Fig.~\ref{figXMCDsimul}(a,b), the sum-rules analysis yields
\rem{$\langle L_z\rangle=0.034$ and
$\langle S_{\rm eff}\rangle=0.026$, corresponding to
$\langle L_z\rangle/\langle S_{\rm eff}\rangle=1.30$.}
\red{$\langle L_z\rangle=0.034(7)$ and
$\langle S_{\rm eff}\rangle=0.024(5)$, corresponding to
$\langle L_z\rangle/\langle S_{\rm eff}\rangle=1.38(15)$. The quoted
uncertainties account for the systematic sensitivity of the sum-rule
analysis to the spectral processing parameters, as detailed in the SM
\cite{Supplementary}.}

\rem{Consequently, the resulting orbital and effective spin magnetic
moments are $m_L=0.034~\mu_{\rm B}$ and
$m_{S_{\rm eff}}=0.052~\mu_{\rm B}$, giving a total effective XMCD
moment $m_{\rm XMCD}^{\rm 300K,6T}
=m_L+m_{S_{\rm eff}}=0.086~\mu_{\rm B}$.}
\red{The corresponding orbital moment is
$m_L=0.034(7)\,\mu_{\rm B}$. For high-spin Co$^{2+}$, the
effective-spin sum rule is subject to a correction associated with
$L_3$--$L_2$ edge mixing and multiplet effects
\cite{Piamonteze2009}. Using the corresponding correction factor
$C_S\simeq1.08$ gives
$m_{S_{\rm eff}}^{\rm corr}=0.052(11)\,\mu_{\rm B}$ and
$m_L/m_{S_{\rm eff}}^{\rm corr}=0.64(7)$. This sizable orbital-to-effective-spin moment ratio is comparable
to values reported experimentally for other high-spin Co$^{2+}$
compounds \cite{Lee_2015,PhysRevB.77.125124,PhysRevB.66.075101}.}

\red{The crystal-field multiplet calculations provide an independent
comparison with the orbital and spin contributions obtained from the
XMCD sum rules. The relevant expectation values were thermally averaged
over the calculated states at $T=300$~K and $\mu_0H=6$~T. In the
$O_h$ limit, the calculation gives
$\langle L_z\rangle/\langle S_z\rangle=1.055$ and
$\langle L_z\rangle/\langle S_{\rm eff}\rangle=1.096$.
Systematic variation of the trigonal crystal-field parameters shows
that the orbital and effective-spin contributions remain of comparable
magnitude throughout the crystal-field regime compatible with the
experimental spectra, while $\langle T_z\rangle$ remains small.
The complete evolution of these quantities with $D_\sigma$ and
$D_\tau$ is presented in the SM \cite{Supplementary}.}

For a direct comparison, vibrating sample magnetometry (VSM) measurements
on an oriented crystal of NaCdCo$_2$F$_7$ were performed in the Physical
Property Measurement System (PPMS14). The magnetic moment measured at
300~K and $\mu_0H=6$~T with field applied along the cubic [100]
direction was found to be
$m^{\rm 300K,6T}_{\rm VSM}=0.092~\mu_{\rm B}$.
\rem{The XMCD-derived moment is therefore in a reasonable agreement with
the bulk magnetization value, differing by approximately $7~\%$. Such a
discrepancy is expected, because the spin sum rule probes the effective
quantity $\langle S_{\rm eff}\rangle =
\langle S_z\rangle+\frac{7}{2}\langle T_z\rangle$, rather than the pure
spin contribution $\langle S_z\rangle$. In NaCdCo$_2$F$_7$, the local
$D_{3d}$ symmetry of the weakly trigonally compressed CoF$_6$ octahedron
is expected to produce a non-zero magnetic dipole term
$\langle T_z\rangle$, introducing an intrinsic uncertainty in the
extracted moment. Multiplet calculations for high-spin Co$^{2+}$
($3d^7$) in an octahedral crystal field of $10Dq=0.9$~eV predict a
finite $\frac{7}{2}\langle T_z\rangle$ contribution of order
$\sim0.05$, assuming atomic $3d$ SOC and full saturation
\cite{Piamonteze2009}. Since the present XMCD measurements were
performed at 300~K and 6~T, where the magnetic polarization remains very
small ($m\approx0.092~\mu_{\rm B}$), the corresponding
$\langle T_z\rangle$ contribution is expected to be substantially
reduced. Additional systematic deviations may arise from the
surface-sensitive nature of the TEY detection mode, including saturation
effects, sample charging, and possible differences between surface and
bulk magnetic response. Consequently, the XMCD-derived moments should be
regarded as semi-quantitative.}
\red{The corrected XMCD-derived moment,
$m_{\rm XMCD}^{\rm 300K,6T}
=m_L+m_{S_{\rm eff}}^{\rm corr}\simeq0.086(13)~\mu_{\rm B}$,
is in good agreement with the bulk magnetization value. The spin sum
rule probes the effective quantity
$\langle S_{\rm eff}\rangle=
\langle S_z\rangle+\frac{7}{2}\langle T_z\rangle$ rather than the pure
spin contribution $\langle S_z\rangle$. The crystal-field multiplet
calculations show that $\langle T_z\rangle$ remains small in the
$O_h$ and weakly trigonal regimes compatible with the experimental
spectra, indicating only a modest magnetic-dipole contribution.
Additional systematic uncertainty may arise from the surface-sensitive TEY detection mode, including saturation effects and
possible differences between the surface and bulk magnetic response. Further details and comparison of the experimental
and calculated moment projections are provided in the SM
\cite{Supplementary}.}

\rem{Despite these limitations, the enhanced
$m_L/m_{S_{\rm eff}}^{\rm corr}=0.64(7)$ ratio
provides clear evidence that the orbital moment of Co$^{2+}$ remains
far from quenched in NaCdCo$_2$F$_7$, highlighting the importance of
SOC in its low-energy magnetic degrees of freedom. In the presence of
the trigonal crystal electric field of the distorted CoF$_6$ octahedron,
such spin-orbit-entangled Co$^{2+}$ moments are expected to give rise
to a strongly anisotropic local $g$-tensor and the local XY-type
anisotropy previously reported in the isostructural
Na$A''$Co$_2$F$_7$ ($A''$ = Ca, Sr) pyrochlores
\cite{Ross2016,Ross2017}.}
\red{The enhanced
$m_L/m_{S_{\rm eff}}^{\rm corr}=0.64(7)$ ratio
provides clear evidence that the orbital moment of Co$^{2+}$ remains
substantially unquenched in NaCdCo$_2$F$_7$, highlighting the
importance of SOC in its low-energy magnetic degrees of freedom.
Such spin-orbit-entangled Co$^{2+}$ moments can exhibit strongly
anisotropic local magnetic properties and are consistent with the
local XY-type anisotropy previously established in the isostructural
Na$A''$Co$_2$F$_7$ ($A''$ = Ca, Sr) pyrochlores
\cite{Ross2016,Ross2017}.} 

We note that due to the insulating nature of NaCdCo$_2$F$_7$, TEY measurements at lower temperatures (10 K and 2 K) suffered from a significantly reduced signal-to-noise ratio due to sample charging effects and reduced electron yield, and were not considered for analysis. Alternative detection modes, including Co $L\alpha$ partial electron yield (PEY) or the O $K\alpha$ inverse partial fluorescence yield (iPFY) mode did not provide sufficient signal quality for quantitative XMCD analysis. Consequently, the determination of orbital and spin magnetic moment was restricted to the (300 K, 6 T) TEY dataset. Under these conditions the paramagnetic moment of the highly frustrated pyrochlore Co$^{2+}$ antiferromagnet remains very low and far from saturation, which limits the quantitative accuracy of the XMCD sum-rules analysis.


\begin{table}[hbtp]
\caption{\red{Crystal-field, spin--orbit coupling (SOC), and atomic
Hamiltonian parameters used for the $O_h$ crystal-field multiplet
calculation of the Co $L_{2,3}$ XAS/XMCD spectra shown in
Fig.~\ref{figXMCDsimul}. The superscripts $i$ and $f$ denote the
initial and final states, respectively. Systematic $C_{3v}$
calculations are presented in the SM \cite{Supplementary}.}}
\setlength{\tabcolsep}{4.0pt}
\renewcommand{\arraystretch}{1.3}

\begin{tabular}{ccccccc}
\hline 
\multicolumn{4}{c||}{\begin{tabular}[c]{@{}c@{}}
Crystal field \\ and SOC \\ ($O_h$)
\end{tabular}}
&
\multicolumn{2}{c|}{\begin{tabular}[c]{@{}c@{}}
Slater--Condon \\ integrals \\ (Hartree--Fock)
\end{tabular}}
&
\begin{tabular}[c]{@{}c@{}}
Scale \\ factor
\end{tabular}
\\ \hline

\multicolumn{4}{c||}{\multirow{3}{*}{$10Dq=0.9$ eV}}
&
\multicolumn{2}{c|}{$F^{2,i}_{dd}=11.605$ eV}
&
\multirow{5}{*}{0.75}
\\

\multicolumn{4}{c||}{}
&
\multicolumn{2}{c|}{$F^{4,i}_{dd}=7.209$ eV}
&
\\

\multicolumn{4}{c||}{}
&
\multicolumn{2}{c|}{$F^{2,f}_{dd}=12.396$ eV}
&
\\ \cline{1-4}

\multicolumn{4}{c||}{$\lambda_{3d}^{i}=0.066$ eV}
&
\multicolumn{2}{c|}{$F^{4,f}_{dd}=7.708$ eV}
&
\\

\multicolumn{4}{c||}{$\lambda_{3d}^{f}=0.083$ eV}
&
\multicolumn{2}{c|}{$F^{2,f}_{pd}=7.260$ eV}
&
\\ \cline{5-7}

\multicolumn{4}{c||}{$\lambda_{2p}^{f}=9.748$ eV}
&
\multicolumn{2}{c|}{$G^{1,f}_{pd}=5.397$ eV}
&
\multirow{2}{*}{0.78}
\\

\multicolumn{4}{c||}{}
&
\multicolumn{2}{c|}{$G^{3,f}_{pd}=3.069$ eV}
&
\\ \hline

\end{tabular}
\label{tabXMCD}
\end{table}

\subsection{Muon spin relaxation spectroscopy}

\begin{figure}[t]
\begin{center}
\includegraphics[scale=1]{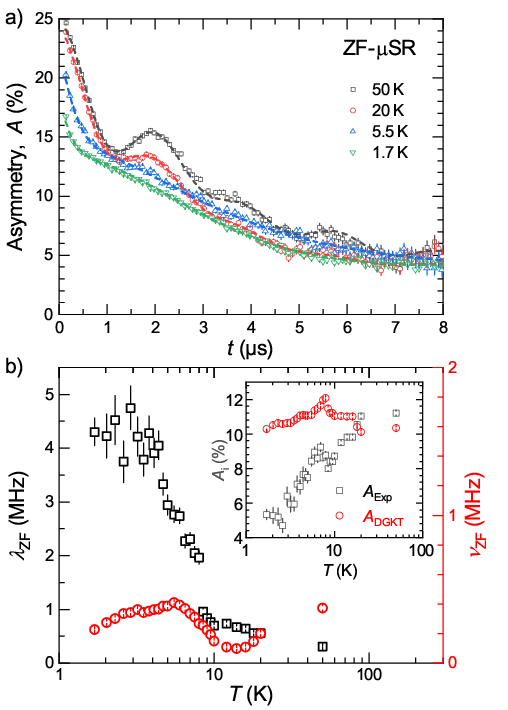}
\caption{(a) Temperature dependence of ZF muon asymmetry. (b) Temperature dependence of fitted ZF muon relaxation rate $\lambda_{\rm ZF}$ and fluctuation rate $\nu_{\rm ZF}$ from Equation \ref{eqA_ZF}, with fitted relaxing amplitudes shown in the inset.} 
\label{fig2} 
\end{center}
\end{figure}

Ground-state spin dynamics were probed by means of muon spin relaxation ($\mu$SR) spectroscopy. Zero-field (ZF) measurements, summarized in Fig. \ref{fig2}(a),  reveal oscillations stemming from the F$^-$--$\mu^+$--F$^-$ state, as previously seen in the isostructural NaCdCu$_2$F$_7$ and NaCaNi$_2$F$_7$ fluoride pyrochlores \cite{Kancko2025,Cai2018} and other fluoride materials \cite{Brewer1986,Wilkinson2020}. At 50 K, the dipolar oscillations are most pronounced and in the cubic lattice, the ZF F-$\mu$-F function $G_{F-\mu-F}(t)$ has the form \cite{Brewer1986} 

\begin{align}
G_{\rm F-\mu-F}(t) = \frac{1}{6} \Big[ \;&
3 + \cos(\sqrt{3}\,\omega_d t + \phi) \nonumber\\
&+ \left( 1-\frac{1}{\sqrt{3}} \right)
\cos\!\left( \frac{3-\sqrt{3}}{2}\,\omega_d t + \phi \right) \nonumber\\
&+ \left( 1+\frac{1}{\sqrt{3}} \right)
\cos\!\left( \frac{3+\sqrt{3}}{2}\,\omega_d t + \phi \right)
\Big],
\label{eq4}
\end{align}

where $\phi$ is a small phase shift to account for a finite time until thermalization of muons in the stopping site, and $\omega_d$ is the dipolar interaction frequency.

As we cool down and approach the spin-glass-like transition, seen by DC and AC susceptibility measurements around $T_f = 4$ K \cite{Kancko2023}, the electronic fluctuations progressively slow down and enter the muon time window, eventually washing out the F$^-$--$\mu^+$--F$^-$ oscillations by the emergence of a broad quasi-static electronic field distribution due to spin freezing. These increasing electronic fluctuations manifest as an exponential damping of the F$^-$--$\mu^+$--F$^-$ signal, $G_{\rm F-\mu-F}(t) e^{-\lambda_{\rm ZF} t}$. 

We note that even at the lowest temperatures, there is an absence of the characteristic minimum then a recovery to a $\frac{1}{3}$-tail at longer times, as expected for a canonical static spin-glass state, following the Kubo-Toyabe relaxation~\cite{KuboToyabe1966}. Instead, a fully monotonic decay of asymmetry is observed at the lowest temperatures, indicating that the local fields are not fully static on the muon timescale, suggesting persistent spin dynamics.  The observation of continued dynamics, in an otherwise spin-glass-like state motivates the use of a Dynamic Gaussian Kubo–Toyabe (DGKT) function \cite{Hayano1979}. This function incorporates an additional relaxation pathway, that takes into account a persistent fluctuating distribution of local magnetic fields. At high fluctuation rate, $\nu$, this collapses to an exponential decay, whilst vanishing fluctuation rate leaves the static Kubo-Toyabe function. At intermediate rates, these fluctuations increasingly wash out the features of the Kubo-Toyabe function, relaxing the expected $\frac{1}{3}$-tail: 

\begin{equation}
    G_{\rm DGKT}(t) = g_{\rm z}(t)e^{-\nu t} + \nu \int_0^t g_{\rm z}(\tau)e^{-\nu \tau} G_{\rm DGKT}(t-\tau) d\tau.
    \label{eq3}
\end{equation}

Here, $\nu$ corresponds to the fluctuation rate of the local fields, and $g_{z}(t)$ in zero field has the form of a static Kubo-Toyabe function with a quasi-static field distribution $\Delta_{\rm ZF}$:
\begin{equation}
g_{\rm z}(t,H=0) = \left[ \frac{1}{3} + \frac{2}{3}(1-\Delta_{\rm ZF}^2 t^2)e^{-\frac{1}{2}\Delta_{\rm ZF}^2 t^2} \right].    
\end{equation}

The full temperature-dependence of the ZF relaxation of asymmetry was found to be best described by a function that combines both the DGKT, and the damped F$^-$--$\mu^+$--F$^-$ oscillations as discrete contributions:

\begin{align}
    A_{\rm ZF}(t) =& A_{\rm exp} G_{\rm F-\mu-F}(t) e^{-\lambda_{\rm ZF} t} \nonumber\\
    &+ A_{\rm DGKT} G_{\mathrm{DGKT}}(t,\Delta_{\rm ZF},\nu_{\rm ZF},H=0) + A_{\rm base} 
    \label{eqA_ZF}
\end{align}

where $A_{\rm exp}, A_{\rm DGKT}$ are the relaxing amplitudes of these components, $\lambda_{\rm ZF}$ is the zero-field muon relaxation rate, and $A_{\rm base}$ is a flat baseline corresponding to non-relaxing muons stopping outside the sample. 

The need for two contributions implies two distinct stopping sites for the muons, which has been seen previously in these materials, and justified with DFT-based muon implantation calculations~\cite{Cai2018}. 

The fits are shown in Fig. \ref{fig2}(a) as dashed lines, and the fitted parameters are summarized in Fig. \ref{fig2}(b). An initial fit was performed at 50 K, where the F-$\mu$-F oscillations are best defined, yielding a phase shift of $\phi = -0.14(3)$ and the dipolar frequency $\omega_d/2\pi = 0.2146(16)$ MHz, corresponding to a $\mu-$F separation $r_{\mu-F} = (\mu_0 \gamma_\mu \gamma_F \hbar / 4\pi \omega_d)^{\frac{1}{3}} = 1.19(1)$ \AA, or equivalently the F-$\mu$-F bond length $2r_{\mu-F} = 2.38(2)$ \AA. Here, $\gamma_\mu/2\pi = 135.539$ MHz/T and $\gamma_F/2\pi = 40.053$ MHz/T are the muon and fluorine gyromagnetic ratios, respectively. These parameters were then fixed and a sequential fit between $1.7 - 50$ K was performed. The flat baseline $A_{\rm base}$ and the quasi-static field distribution $\Delta_{\rm ZF}$ proved to be mostly temperature-independent, and were subsequently fixed to $A_{\rm base} = 3.2 ~\%$ and $\Delta_{\rm ZF} = 0.3$ MHz, corresponding to a field distribution of $\delta H_{\rm loc} = \Delta_{\rm ZF}/\gamma_\mu = 22.1$ G. The temperature dependence of the zero-field relaxation rate $\lambda_{\rm ZF}$, and the field fluctuation rate $\nu_{\rm ZF}$, can be seen in Fig. \ref{fig2}(b). We note a steep increase in $\lambda_{\rm ZF}$ below 10 K, reaching up to $\sim\!5$ MHz. This is evidently due to the fact that the extremely fast relaxation of asymmetry at short times, caused by the large Co$^{2+}$ magnetic moment of $\mu_{\rm eff} = 5.4 ~\mu_{\rm B}$, cannot be fully resolved with the limited time resolution of the MUSR pulsed-source instrument. This also explains the loss of relaxing amplitude $A_{\rm exp}$, seen in the inset of Fig. $\ref{fig2}$(b), with decreasing temperature. The fluctuation rate $\nu_{\rm ZF}$ shows a peak centered around 6 K, reaching $\nu_{\rm ZF} \sim 0.4$ MHz, indicating increased dynamics during the spin-freezing process. 

\begin{figure}[t]
\begin{center}
\includegraphics[scale=1]{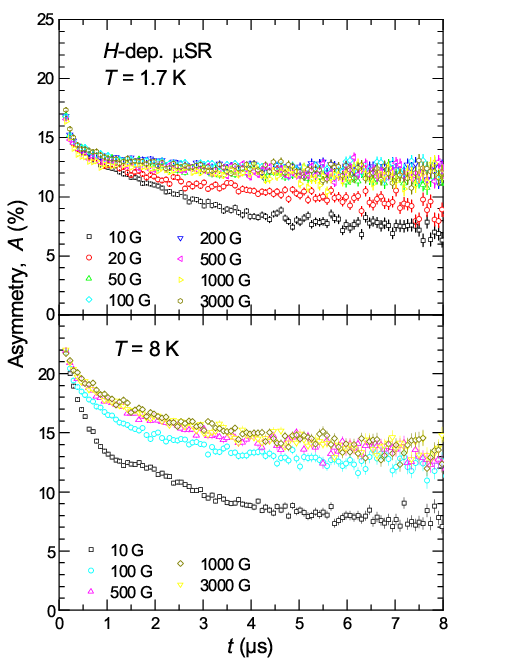}
\caption{Field dependence of muon asymmetry  at $T=1.7$ K (upper panel) and  $T=8$ K (lower panel).} 
\label{fig3} 
\end{center}
\end{figure}

To compare the dynamics below and above the spin-freezing transition expected at $T_f = 4$ K, we measured the field dependence of muon asymmetry at $T=1.7$ and 8 K, seen in Fig. \ref{fig3}. The rapid loss of asymmetry at $t<0.5 ~\mu$s is evident at 1.7 K, followed by a slow relaxation which saturates at 13 \% in an applied field of just $\mu_0H_{\rm LF} =$ 100 G. This suggests the presence of quasi-static internal fields with a dynamically fluctuating component. At 8 K, above $T_f$, we see an exponential relaxation confirming a fast-fluctuating paramagnetic regime. 

\begin{figure}[t]
\begin{center}
\includegraphics[scale=1]{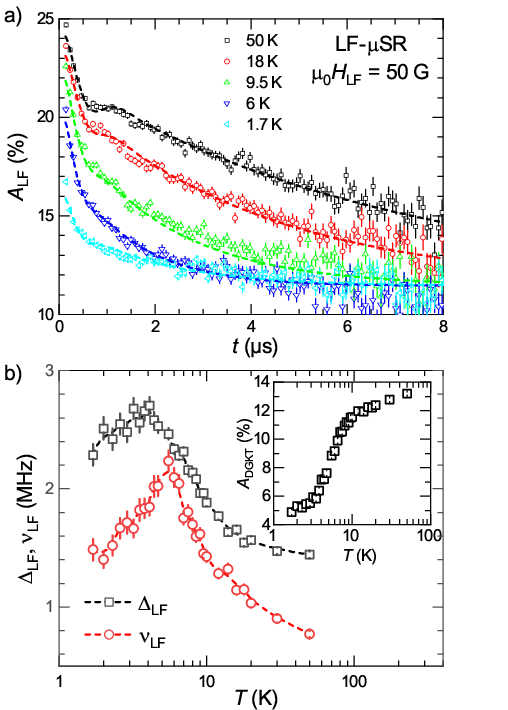}
\caption{(a) Temperature dependence of LF 50 G muon asymmetry. (b) Temperature dependence of fitted quasistatic field distribution $\Delta_{\rm LF}$ and fluctuation rate $\nu_{\rm LF}$ from Equation \ref{eqA_LF}, with the relaxing amplitude $T$-dependence shown in the inset.} 
\label{fig4} 
\end{center}
\end{figure}

To quantitatively describe the spin dynamics in this system, the interaction within the F-$\mu$-F state could be decoupled using a 50 G longitudinal field (LF), leaving the DGKT response dominated by electronic fluctuations. The LF 50 G temperature dependence can be seen in Fig. \ref{fig4}(a). Again, we note a rapid depolarization at $t<0.5 ~\mu$s, followed by a slow relaxation of asymmetry, with a small residual F-$\mu$-F hump around 1 $\mu$s visible at high temperatures. The data were fitted using the function
\begin{equation}
    A_{\rm LF}(t) = A_{\rm DGKT}G_{\rm DGKT}(t,\Delta_{\rm LF}, \nu_{\rm LF}, H_{\mathrm{LF}}=50 ~\mathrm{G}) + A_{\rm base} 
    \label{eqA_LF}
\end{equation}

where $G_{\rm DGKT}(t,\Delta_{\rm LF}, \nu_{\rm LF}, H_{\mathrm{LF}}=50 ~\mathrm{G})$ is defined the same way as in Equation \ref{eqA_ZF}, but uses a
field-dependent $g_z(t,H_{\rm LF})$ inside an applied field $\mu_0 H_{\rm LF} = \omega_0/(2\pi \gamma_\mu)$ as

\begin{align}
  g_{z}(t, H_{\rm LF}) = \Big[ 1 &- 2\frac{\Delta^2}{\omega_0^2} \left( 1- \cos(\omega_0 t) e^{-\frac{1}{2}\Delta_{\rm LF}^2t^2} \right)  \nonumber \\ &+ 2\frac{\Delta^4}{\omega_0^4}\omega_0 \int_0^t \sin(\omega_0 \tau) e^{-\frac{1}{2}\Delta_{\rm LF}^2 \tau^2} d\tau \Big].
\end{align}

The fits can be seen in Fig. \ref{fig4}(a) as dashed lines. The flat background remained largely temperature-independent, and was fixed to $A_{\rm base} = 11.4\%$, now corresponding to muons stopping outside the sample as well as muons in quasi-static local field environments easily repolarized by the weak longitudinal field. The sequentially-fitted parameters are summarized in Fig. \ref{fig4}(b). We observe a peak in the quasi-static field distribution $\Delta_{\rm LF}$ centered around 4 K, as well as a peak in the fluctuation rate $\nu_{\rm LF}$ at 6 K, consistent with a critical slowing down of spin fluctuations near the freezing transition. The maximum value of $\Delta_{\rm LF} = 2.70(7)$ MHz corresponds to the quasi-static local field distribution $\delta H_{\rm loc} = \Delta_{\rm LF}/\gamma_\mu = 199(5)$ G. The relaxing amplitude, seen in the inset of Fig. \ref{fig4}(b), drops similarly as in the ZF measurement, hinting at the difficulty of fitting the unresolved fast-relaxing part of asymmetry below $t < 0.5 ~\mu$s. Together, these results indicate a partially frozen magnetic state with persistent low-temperature spin dynamics, rather than a canonical static spin glass. Similar partial freezing and continued spin dynamics was seen in the isostructural analogue NaCaNi$_2$F$_7$ \cite{Cai2018, Plumb2019}.

\subsection{Nuclear magnetic resonance}

\begin{figure}[htbp]
\begin{center}
\includegraphics[scale=1]{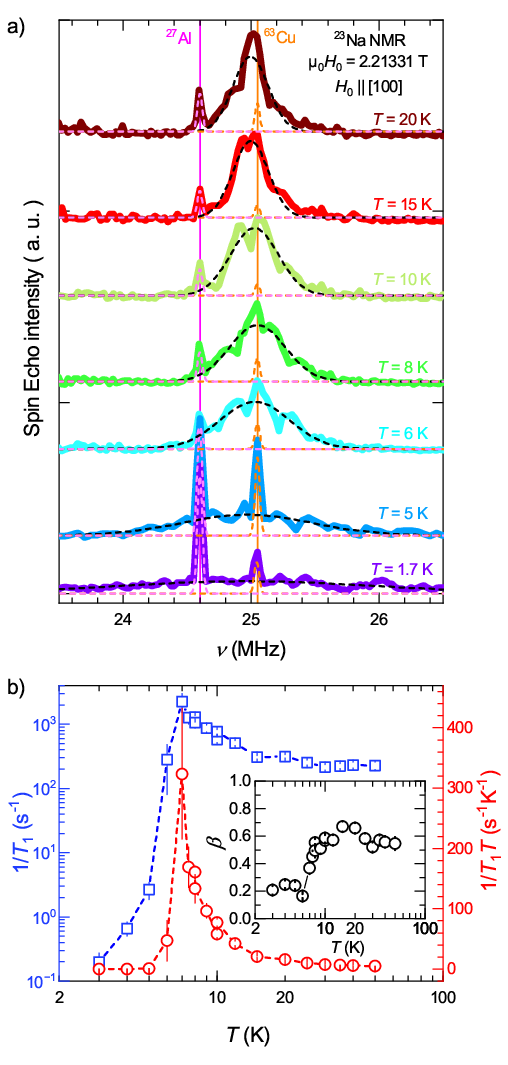}
\caption{(a) Frequency-swept $^{23}$Na NMR spectra at selected temperatures measured in a constant magnetic field of $\mu_0H_0=2.21331$~T applied along the [100] direction. For clarity, each spectrum is normalized by its maximum intensity. In addition to the intrinsic $^{23}$Na signal, extrinsic signals arising from the $^{27}$Al NMR of the Al foil used as a field marker and from a small $^{63}$Cu NMR contribution associated with the copper soldering pad are also indicated. (b) Temperature dependence of $1/T_1$ determined from stretched-exponential fits to the nuclear magnetization recovery curves measured at the $^{23}$Na NMR peak. The corresponding $1/T_1T$, obtained by dividing $1/T_1$ by temperature, is plotted on the right axis. The inset shows the temperature dependence of the stretching exponent $\beta$ obtained from the fits.} 
\label{fig5} 
\end{center}
\end{figure}

Temperature-dependent $^{23}$Na NMR spectra measured between $1.7-20$ K on an oriented single crystal in an applied field $\mu_0 H_0 = 2.21331$ T parallel to the [100] direction are shown in Fig. \ref{fig5}(a).
The spectra consist of three sets of resonance lines: (i) the intrinsic $^{23}$Na signal from the sample ($I = 3/2$, $\gamma_{\rm Na}/2\pi = 11.2625$ MHz/T, 100 \% abundance), (ii) an extrinsic $^{27}$Al signal ($I=5/2$, $\gamma_{\rm Al}/2\pi = 11.094$ MHz/T, 100 \% abundance) originating from a small piece of pure Al foil mounted adjacent to the sample as a field marker, and (iii) an extrinsic $^{63}$Cu signal ($I=3/2$, $\gamma_{\rm Cu}/2\pi = 11.285$ MHz/T, 69.09 \% abundance) arising from a copper contact pad on the sample holder. The weak $^{63}$Cu signal is too small to be resolved at higher temperatures and becomes discernible only below $\sim$5 K in the present spectra.

In the pyrochlore structure, the $^{23}$Na nuclei occupy the axially symmetric $A$ site (point symmetry $\overline{3}m$) where they are randomly distributed with Cd$^{2+}$ ions with equal probability \cite{Kancko2023}. This site symmetry results in an axially symmetric electric field gradient (EFG) with the asymmetry parameter $\eta=0$, whose principal axis is oriented along the local $\langle 111 \rangle$ directions. For an applied field along $H_0 \parallel [100]$, the angle between the field and the EFG principal axis corresponds to the magic angle $\theta = 54.7 \degree$, causing the quadrupolar shift as a first-order perturbation proportional to $[3 \cos^2(\theta)-1]$ to vanish. As a result, only a single central transition is observed, consistent with the previous $^{23}$Na NMR study on the isostructural NaCaCo$_2$F$_7$ \cite{Sarkar2017}. 

Correspondingly, the spectra seen in Fig, \ref{fig5}(a) were decomposed into three Gaussian components. The positions and linewidths of the extrinsic $^{27}$Al and $^{63}$Cu signals were fixed to their well-defined values at 1.7 K, while their amplitudes and the parameters of the $^{23}$Na line were allowed to vary. At high temperatures (10–20 K), the $^{23}$Na line is relatively narrow, consistent with a paramagnetic state characterized by fast electronic spin fluctuations that average out local hyperfine fields.
Upon cooling below 10 K, we observe a continual broadening of the $^{23}$Na central transition peak, reflecting a development of a spatially inhomogeneous distribution of quasi-static internal fields at the $^{23}$Na site.
This broad field distribution reflects the spatial randomness of the local magnetic environment, as expected when Co$^{2+}$ spin dynamics slow down and short-range magnetic correlations become increasingly inhomogeneous.
In such a regime, the same magnetic randomness is also expected to enhance transverse relaxation through slowly fluctuating local fields and a distribution of local relaxation rates.
Therefore, the strong reduction of the observable $^{23}$Na signal intensity below $\sim$5 K can be naturally understood as a wipeout-like effect, in which a fraction of the nuclei relaxes too rapidly in the transverse channel for the spin echo to be detected within the experimental time window.
At 1.7 K, the fitted full width at half maximum (FWHM) of the $^{23}$Na resonance is $\Delta \nu_{\rm Na}=2.4(1)$ MHz, corresponding to a local field distribution $\delta H_{\rm loc} = \Delta \nu_{\rm Na}/\gamma_{\rm Na} = 2130(90)$ G at the $^{23}$Na site.

The nuclear spin-lattice relaxation rate $1/T_1$ was determined at each temperature by measuring the recovery of the nuclear magnetization at the position of the resonance peak.
The recovery curves were fitted using a stretched exponential function:

\begin{align}
    \frac{M(\infty)-M(t)}{M(\infty)} &= a\exp \{-\left( t/T_1 \right)^{\beta} \}.
    \label{eq6}
\end{align}


Here, $M(t)$ is the nuclear magnetization at time $t$ after the saturation pulse and $M(\infty)$ its equilibrium value.
The coefficient $a$ accounts for the inversion or saturation efficiency of the nuclear magnetization produced by the pulse sequence.
When the stretching exponent $\beta=1$, there is a single uniform relaxation time, and when $\beta<1$ this indicates a distribution of inhomogeneous dynamics.

The nuclear spin-lattice relaxation rate $1/T_1$, shown on the left axis in Fig. \ref{fig5}(b), exhibits approximately a 10-fold increase upon cooling from 60 K to 8 K, followed by an abrupt drop by four orders of magnitude upon cooling to 3 K. This is in contrast to canonical spin glasses, where a peak in $1/T_1$ is typically observed due to critical slowing down.
The quantity $1/T_1T$, which probes the low-frequency limit of the momentum \textbf{q}-averaged dynamical spin susceptibility as 

\begin{equation}
 (1/T_1T) \sim \sum\limits_\textbf{q} |A_\textbf{q}|^2 \left( \chi''(\textbf{q},\omega_0)/\omega_0 \right) _{\omega_0 \rightarrow 0}   
\end{equation} 

shows an approximately Curie-like behavior with a maximum at 7 K.
The rapid decrease below this temperature suggests that the dominant spectral weight of the spin fluctuations shifts to frequencies lower than the NMR window ($< ~\sim \!10^7$ Hz) as the system enters the freezing regime.
It should be noted, however, that in this temperature range the observable $^{23}$Na signal intensity is also strongly reduced. Therefore, the measured $1/T_1$ should be regarded as an effective relaxation rate of the remaining observable NMR component, while nuclei experiencing extremely fast transverse relaxation may be lost from the detected echo signal.

In contrast to the isostructural NaCaCo$_2$F$_7$ \cite{Sarkar2017}, where a well-defined Bloembergen–Purcell–Pound (BPP) peak in $1/T_1(T)$ was observed \cite{Bloembergen1948} as

\begin{equation}
    1/T_1(T) = \gamma^2 h_0^2 \frac{\tau_C(T)}{1+\omega_0^2\tau_C(T)^2},
\end{equation}

associated with thermally activated low-energy spin fluctuations with a correlation time $\tau_C(T)=\tau_0\exp(E_a/T)$, no such peak is detected in NaCdCo$_2$F$_7$. 
While a BPP maximum typically arises when a single characteristic correlation time $1/\tau_C(T)$ crosses the NMR resonance frequency $\omega_0$, our data indicate a broad distribution of fluctuation rates, as evidenced by the strongly reduced stretching exponent $\beta \sim 0.2$ below 7 K, as seen in the inset of Fig. \ref{fig5}(b).
In this situation, different regions of the sample satisfy the BPP condition at different temperatures, leading to a broadening or masking of the expected peak.
The absence of a sharp BPP maximum therefore points to a highly inhomogeneous slowing down of spin dynamics, consistent with strong bond disorder on the pyrochlore lattice.
This picture is also compatible with $\mu$SR results suggesting persistent low-temperature spin fluctuations.

\section{Discussion}

The combined high-field magnetization and XMCD results establish
NaCdCo$_2$F$_7$ as a spin-orbit-entangled pyrochlore antiferromagnet
whose low-energy magnetic properties are governed by a well-isolated
$J_{\rm eff}=\frac{1}{2}$ Kramers doublet. Although direct determination
of the crystal-field level scheme by INS is prohibited by the strong
neutron absorption of natural Cd, the large effective $g$ factor of 6.56
extracted from high-field magnetization, together with
\rem{the substantial orbital contribution
$\langle L_z \rangle / \langle S_{\rm eff} \rangle = 1.30$ at 300 K
revealed by XMCD,}
\red{the substantial unquenched orbital contribution revealed by XMCD,
$m_L/m_{S_{\rm eff}}^{\rm corr}=0.64(7)$,}
provide strong evidence for the importance of SOC in the Co$^{2+}$
ground state. These observations are fully consistent with previous
studies of the isostructural NaCaCo$_2$F$_7$ and NaSrCo$_2$F$_7$
pyrochlores, where a strongly anisotropic local $XY$-type $g$ tensor and
a spin-orbit-entangled $J_{\rm eff}=\frac{1}{2}$ Kramers doublet have
been established by INS experiments. 

The Co $L_{2,3}$ XAS spectra are found to be \rem{dominated by the octahedral crystal field, with only a weak trigonal
distortion required to reproduce the experimental line shape, indicating
that the local electronic structure remains primarily governed by the
cubic ligand environment.} \red{consistent with a predominantly octahedral ($O_h$) crystal field,
while remaining compatible with weak trigonal perturbations.}
This observation is consistent with recent RIXS measurements on
NaCaNi$_2$F$_7$, which demonstrated that the local transition-metal
electronic structure of fluoride pyrochlores is remarkably robust against
$A$-site disorder and remains close to ideal octahedral symmetry despite
the presence of significant Na/Ca site mixing \cite{DiScala2026}. The
dominant consequence of the cationic $A$-site mixing therefore does not
seem to lie in changing the local electronic structure, but rather in
producing randomness in the magnetic exchange pathways, resulting in bond
disorder on the pyrochlore lattice.

The low-temperature $\mu$SR and $^{23}$Na NMR results provide complementary evidence for how this bond disorder manifests itself in the spin dynamics. Both techniques reveal a progressive slowing down of fluctuations below approximately 10 K, consistent with the development of increasingly inhomogeneous short-range correlations. In $\mu$SR, this appears as a strong enhancement of the muon relaxation rate and the emergence of quasi-static and fluctuating local fields. Meanwhile, NMR detects a pronounced broadening of the $^{23}$Na line and a loss of intensity in a wipeout-like effect, together with a stretched-exponential relaxation of nuclear magnetization characterized by a strongly suppressed stretching exponent $\beta \sim 0.2$ below 8 K. The absence of a conventional static Kubo--Toyabe recovery in $\mu$SR, together with the suppression of a well-defined BPP peak in $1/T_1(T)$ previously seen in NaCaCo$_2$F$_7$ \cite{Sarkar2017}, indicate that spin freezing proceeds through a broad distribution of local fluctuation rates rather than through a uniform slowing down process. Such behavior points to a highly inhomogeneous magnetic state arising from bond disorder.

Importantly, the low-temperature state deviates significantly from a canonical spin glass. While DC and AC susceptibility measurements identify a bulk spin-glass-like transition at $T_f \sim 4$ K \cite{Kancko2023}, both local-probe techniques, $\mu$SR and $^{23}$Na NMR, reveal a more gradual slowing down of spin fluctuations beginning already below $\sim 8$ K, indicating that the freezing process develops continuously over an extended temperature range. 
Despite the apparent spin freezing, persistent low-temperature spin dynamics remain clearly visible on the muon timescale. Such coexistence of quasi-static and fluctuating spin components closely resembles the behavior reported for the isostructural $S=1$ Heisenberg pyrochlore antiferromagnet NaCaNi$_2$F$_7$ \cite{Cai2018}, where frozen magnetism similarly coexists with continued dynamics at low temperature. The present results therefore support an emerging picture in which bond-disordered Na$A''M_2$F$_7$ ($A''$ = Ca, Sr, Cd; $M$ = Co, Ni, Mn, Fe) fluoride pyrochlores frequently realize unconventional spin-glass-like states, where geometric frustration inhibits complete static freezing even well below $T_f$.

A striking comparison with NaCdCu$_2$F$_7$ arises \cite{Kancko2025}, as two fluoride pyrochlores with comparable bond disorder realize fundamentally different ground states: a random-singlet state in the  Heisenberg $S=\frac{1}{2}$ Cu$^{2+}$ system, and an unconventional spin-glass-like state in the spin-orbit-entangled $J_{\rm eff}=\frac{1}{2}$ Co$^{2+}$ analogue. While NaCdCu$_2$F$_7$ exhibits a disorder-driven random-singlet state characterized by power-law scaling of bulk and local-probe thermodynamic quantities, with data collapse onto universal curves and the absence of bulk spin freezing, NaCdCo$_2$F$_7$ develops a partially-frozen spin-glass-like state with persistent low-temperature spin dynamics. This contrast is particularly remarkable because both materials host (effective) spin-$\frac{1}{2}$ magnetic moments on the geometrically frustrated pyrochlore lattice, where strong quantum fluctuations would naively be expected to suppress magnetic freezing. However, XMCD measurements demonstrate that the Co$^{2+}$ moments retain a substantial unquenched orbital contribution, \red{ highlighting the importance
of strong spin-orbit coupling.} \rem{and strong spin-orbit coupling, giving rise to a much larger magnetic moment together with local $XY$ anisotropy and anisotropic exchange interactions.} \red{Together with the local XY anisotropy and
anisotropic exchange interactions established in the isostructural
Na$A''$Co$_2$F$_7$ ($A''=$ Ca, Sr) analogues, these} \rem{These} ingredients appear to fundamentally alter the response to bond disorder and favor glassy freezing over random-singlet formation. This comparison highlights spin-orbit coupling as a key tuning parameter governing the fate of bond-disordered pyrochlore antiferromagnets.

\section{Conclusions}
 
We have investigated the magnetic ground state of the bond-disordered pyrochlore antiferromagnet NaCdCo$_2$F$_7$ using high-field magnetization, Co $L_{2,3}$-edge XAS/XMCD, $\mu$SR, and $^{23}$Na NMR measurements. 
High-field magnetization supports the presence of a well-isolated spin-orbit-entangled $J_{\rm eff}=\frac{1}{2}$ ground-state Kramers doublet characterized by a large effective $g$ factor and strong antiferromagnetic interactions. XMCD measurements reveal a substantial unquenched orbital contribution to the Co$^{2+}$ magnetic moment, consistent with strong SOC and the emergence of local $XY$-type single-ion anisotropy seen in the isostructural Na$A''$Co$_2$F$_7$ ($A''$ = Ca, Sr) analogues. Both $\mu$SR and NMR measurements demonstrate a progressive slowing down of spin fluctuations below approximately 10 K, culminating in a partially frozen magnetic state. However, persistent low-temperature spin dynamics are observed by $\mu$SR, indicating that the frozen state deviates from a conventional static spin glass. 
A comparison with the isostructural $S=\frac{1}{2}$ Heisenberg antiferromagnet NaCdCu$_2$F$_7$ with a random singlet ground state reveals a fundamentally different response of spin-orbit-entangled $J_{\rm eff} = \frac{1}{2}$ Co$^{2+}$ moments to bond disorder on the pyrochlore lattice. Together, these results establish NaCdCo$_2$F$_7$ as a disorder-driven spin-glass-like pyrochlore exhibiting persistent low-temperature dynamics arising from strongly spin-orbit-coupled $J_{\rm eff}=\frac{1}{2}$ moments.

\begin{acknowledgments}
We acknowledge funding from Charles University in Prague, within Grant Agency Univerzita Karlova program (GAUK 48924). Crystal growth, structural analysis and magnetic properties measurements were carried out in the MGML (http://mgml.eu/), which is supported within the program of Czech Research Infrastructures (project no. LM2023065). This work was supported by the Ministry of Education, Youth and Sports of the Czech Republic through the INTER-EXCELLENCE II program (LUABA24056), by the Grant Agency of the Czech Republic (26-23051S), as well as by the HLD at HZDR, member of the European Magnetic Field Laboratory (EMFL) and by EPSRC (UK) via its membership to the EMFL (grant no. EP/N01085X/1).
H. S. and Y. T. acknowledge support from Japan Society for the Promotion of Science (JSPS) through KAKENHI (JP24KK0062 and JP23K25829).
Work at the Japan Atomic Energy Agency (JAEA) was partially supported by the JAEA REIMEI Research Program. XMCD experiments were performed at the BOREAS BL-29 beamline of the ALBA Synchrotron Light Facility (proposal no. 2024098632) with the collaboration of ALBA staff. The research leading to this result has been co-funded by the project NEPHEWS under the Grant Agreement No 101131414 from the EU Framework Programme for Research and Innovation Horizon Europe. 
The authors additionally thank Gael Bastien for insightful discussions, and Martin Misek for performing complementary magnetization measurements.
\end{acknowledgments}

\bibliography{bibliography}

\end{document}


\title{Supplemental materials: Spin-orbit-entangled
$J_{\rm eff}=\frac{1}{2}$ magnetism and unconventional spin freezing
in the bond-disordered pyrochlore antiferromagnet NaCdCo$_2$F$_7$}

\author{A. Kancko}
\affiliation{Charles University, Faculty of Mathematics and Physics,
Department of Condensed Matter Physics, Ke Karlovu 5, 121 16, Prague,
Czech Republic}

\author{H. Sakai}
\affiliation{Advanced Science Research Center, Japan Atomic Energy Agency,
Tokai, Ibaraki 319-1195, Japan}

\author{J. Herrero-Mart\'in}
\affiliation{ALBA Synchrotron Light Source, Carrer de la Llum 2-26,
08290 Cerdanyola del Vall\`es, Barcelona, Spain}

\author{A. Berlie}
\affiliation{ISIS Neutron and Muon Source, Rutherford Appleton Laboratory,
Science and Technology Facilities Council, Chilton,
Oxfordshire OX11 0QX, U.K.}

\author{M. Uhlarz}
\affiliation{Hochfeld-Magnetlabor Dresden (HLD-EMFL),
Helmholtz-Zentrum Dresden-Rossendorf (HZDR),
01328 Dresden, Germany}

\author{T. Haidamak}
\affiliation{Charles University, Faculty of Mathematics and Physics,
Department of Condensed Matter Physics, Ke Karlovu 5, 121 16, Prague,
Czech Republic}

\author{M. Klicpera}
\affiliation{Charles University, Faculty of Mathematics and Physics, Department of Condensed Matter Physics, Ke Karlovu 5, 121 16, Prague, Czech Republic}

\author{Y. Tokunaga}
\affiliation{Advanced Science Research Center, Japan Atomic Energy Agency,
Tokai, Ibaraki 319-1195, Japan}

\author{R. H. Colman}
\affiliation{Charles University, Faculty of Mathematics and Physics,
Department of Condensed Matter Physics, Ke Karlovu 5, 121 16, Prague,
Czech Republic}

\date{\today}

\maketitle

\section{C\MakeLowercase{o}$^{2+}$ $L_{2,3}$-edge XAS \& XMCD simulations}

\subsection{Sensitivity to the trigonal crystal field}

To examine the sensitivity of the calculated Co $L_{2,3}$-edge XAS and
XMCD spectra to a weak trigonal distortion of the octahedral crystal
field, additional crystal-field multiplet calculations were performed using
Crispy \cite{retegan_crispy}, a graphical interface to the Quanty
computational framework \cite{quanty}. These calculations are not
treated as fits to the experimental spectra. Rather, the crystal-field
parameters are varied systematically and the resulting spectra are
compared with experiment to identify regimes consistent with the
observed spectral line shapes. The resulting crystal-field parameters
should therefore be regarded as qualitative rather than as a
quantitative determination of the crystal-field splitting.

To quantify the agreement between calculated and experimental spectra,
in addition to inspection of the spectral line shapes, we define the
normalized residual factor

\begin{equation}
    R =
    \left[
    \frac{
    \displaystyle\int_{E_1}^{E_2}
    \left[
    I^{\rm exp}(E)-I^{\rm calc}(E)
    \right]^2\,dE
    }{
    \displaystyle\int_{E_1}^{E_2}
    \left[
    I^{\rm exp}(E)
    \right]^2\,dE
    }
    \right]^{1/2},
    \label{eq:Rfactor}
\end{equation}

where $I^{\rm exp}(E)$ and $I^{\rm calc}(E)$ are the experimental and
calculated intensities, respectively, and $E_1$ and $E_2$ define the
common energy interval. The residual factors were evaluated separately
for the XAS and XMCD spectra over the same energy interval for all
calculations. Smaller values of $R$ indicate closer agreement with
experiment. Since the model parameters were not optimized by a fitting
procedure and experimental uncertainties are not included in
Eq.~\ref{eq:Rfactor}, $R$ is used only as a relative measure for
comparing the calculated spectra.

The calculations were performed for the Co$^{2+} L_{2,3}$ edge in $C_{3v}$ symmetry at
$T=300$~K and $\mu_0H=6$~T, corresponding to the experimental
conditions. The octahedral crystal-field splitting was fixed at
$10Dq=0.9$~eV, the Slater-integral reduction factors at $F_k=0.75$ and
$G_k=0.78$, and the spin-orbit coupling scaling factor at $\zeta=1$.
The calculated spectra were broadened using Lorentzian and Gaussian
functions with full widths at half maximum (FWHM) of $0.43$~eV and
$0.10$~eV, respectively. To isolate the effects of the two trigonal
crystal-field parameters, $D_\sigma$ and $D_\tau$ were varied
separately. We first fixed $D_\tau=0$ and varied $D_\sigma$ from
$-0.15$ to $+0.15$~eV in steps of $0.05$~eV. For
$D_\sigma=D_\tau=0$, the trigonal contribution vanishes and the
calculation reduces to the octahedral $O_h$ crystal-field limit.
Simultaneous variations of $D_\sigma$ and $D_\tau$ were not explored,
since the purpose of these calculations is to establish the sensitivity
of the spectra to trigonal perturbations rather than to determine a
unique set of crystal-field parameters.

\begin{figure}[htbp]
    \centering
    \includegraphics[width=\linewidth]{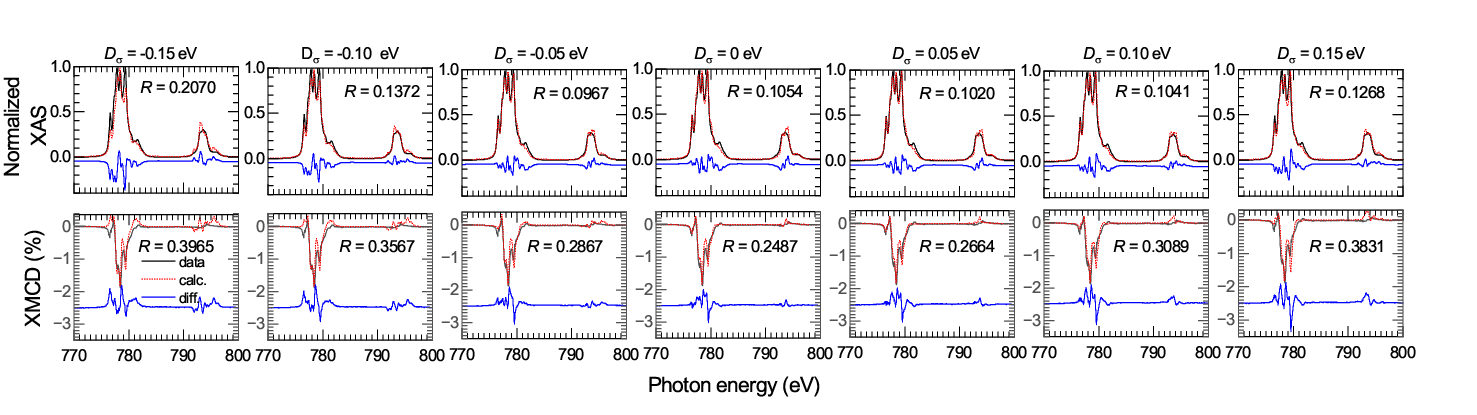}
    \caption{
    Comparison of the experimental Co $L_{2,3}$-edge XAS and XMCD
    spectra with the corresponding crystal-field multiplet calculations for
    different values of $D_\sigma$ at fixed $D_\tau=0$. Black solid
    lines denote the experimental data, red dotted lines the calculated
    spectra, and blue solid lines the difference between calculation
    and experiment. The normalized residual factors $R$, defined by
    Eq.~\ref{eq:Rfactor}, are indicated for each spectrum.
    }
    \label{figS1}
\end{figure}

Figure~\ref{figS1} compares the calculated spectra with the experimental
XAS and XMCD data. The XAS line shape is relatively insensitive to weak
variations of $D_\sigma$, whereas the XMCD response exhibits a stronger
dependence on the trigonal crystal field. For XAS, the smallest residual
is obtained for $D_\sigma=-0.05$~eV ($R=0.0967$), compared with
$R=0.1054$ in the $O_h$ limit and $R=0.1020$ for
$D_\sigma=+0.05$~eV. The differences are small, and all three
calculations give similar descriptions of the experimental XAS line
shape.

The XMCD residual shows a clearer minimum at the $O_h$ limit, with
$R=0.2487$ for $D_\sigma=0$, compared with 0.2867 and 0.2664 for
$D_\sigma=-0.05$ and $+0.05$~eV, respectively. Larger
$|D_\sigma|$ values produce progressively larger XMCD residuals,
reaching $R=0.3965$ and 0.3831 for $D_\sigma=-0.15$ and
$+0.15$~eV, respectively. The residual analysis therefore favors a
crystal field close to the $O_h$ limit, although weak trigonal
perturbations cannot be excluded and a unique value of $D_\sigma$
cannot be determined from the present results.

\begin{figure}[htbp]
    \centering
    \includegraphics[width=\linewidth]{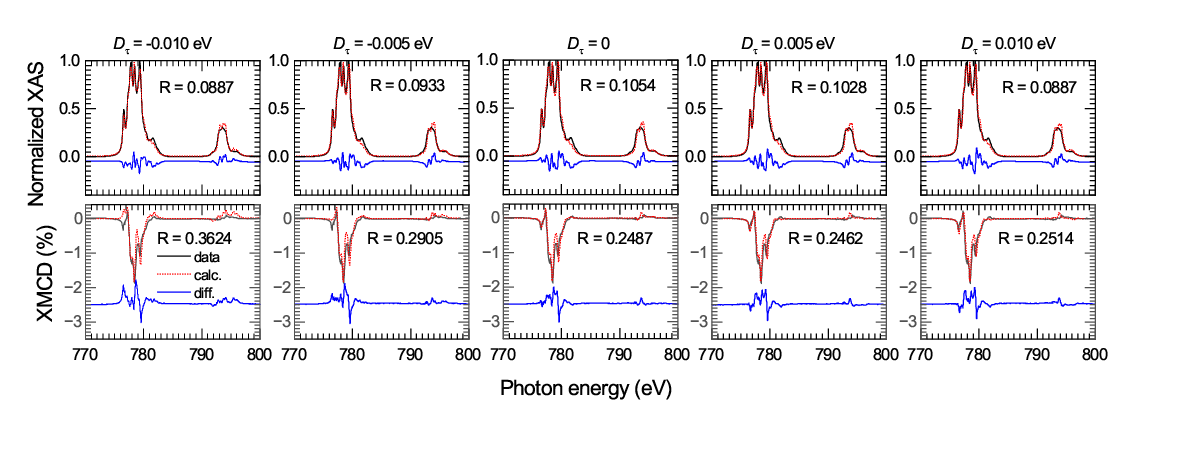}
    \caption{
    Comparison of the experimental Co $L_{2,3}$-edge XAS and XMCD
    spectra with the corresponding crystal-field multiplet calculations for
    different values of $D_\tau$ at fixed $D_\sigma=0$. Black solid
    lines denote the experimental data, red dotted lines the calculated
    spectra, and blue solid lines the difference between calculation
    and experiment. The normalized residual factors $R$, defined by
    Eq.~\ref{eq:Rfactor}, are indicated for each spectrum.
    }
    \label{figS2}
\end{figure}

The sensitivity to the second trigonal crystal-field parameter was
examined independently by fixing $D_\sigma=0$ and varying $D_\tau$
from $-0.010$ to $+0.010$~eV. As shown in Fig.~\ref{figS2}, even
small changes in $D_\tau$ modify the calculated spectra, particularly
the XMCD response. For XAS, the residual decreases from $R=0.1054$
in the $O_h$ limit to 0.0933 for $D_\tau=-0.005$~eV and 0.1028
for $D_\tau=+0.005$~eV, with the smallest value, $R=0.0887$,
obtained at both $D_\tau=-0.010$ and $+0.010$~eV. The XAS residual
therefore does not identify a preferred sign of $D_\tau$ within the
range considered.

For XMCD, negative $D_\tau$ increases the residual substantially, from
$R=0.2487$ in the $O_h$ limit to 0.2905 and 0.3624 for
$D_\tau=-0.005$ and $-0.010$~eV, respectively. Positive $D_\tau$
produces only small changes, with $R=0.2462$ and 0.2514 for
$D_\tau=+0.005$ and $+0.010$~eV, respectively. Although the lowest
XMCD residual occurs at $D_\tau=+0.005$~eV, its difference from the
$O_h$ value is marginal and does not provide evidence for a finite
$D_\tau$.

Taken together, the $D_\sigma$ and $D_\tau$ series favor a crystal
field close to the $O_h$ limit. The XMCD residual is minimized at
$D_\sigma=0$, while variation of $D_\tau$ yields no significant
improvement over the $O_h$ result and the XAS and XMCD residuals do
not select a common finite $D_\tau$. Weak trigonal perturbations remain
compatible with the spectra, but their magnitude and sign cannot be
determined uniquely from the present results. We therefore retain the
$O_h$ calculation as the minimal description in the main text.
Additional constraints from the calculated orbital and spin expectation
values are considered below.

\FloatBarrier

\subsection{Calculation of orbital and spin expectation values}

In addition to the calculated spectral line shapes, the crystal-field multiplet
calculations provide the expectation values of the orbital, spin, total
angular momentum, and magnetic-dipole operators for each eigenstate.
Since the calculations correspond to $T=300$~K and $\mu_0H=6$~T,
several states are thermally populated. The quantities discussed below
are therefore thermal averages over the calculated eigenstates,

\begin{equation}
    \langle A\rangle_T
    =
    \sum_i dZ_i \langle A\rangle_i ,
    \label{eq:thermal_average}
\end{equation}

where $dZ_i$ is the normalized Boltzmann weight of the $i$th state
reported by Crispy. The same averaging procedure was applied to
$\langle L^2\rangle$, $\langle S^2\rangle$, $\langle J^2\rangle$,
$\langle L_z\rangle$, $\langle S_z\rangle$, $\langle J_z\rangle$,
$\langle T_z\rangle$, and
$\langle\mathbf{l}\cdot\mathbf{s}\rangle$, where
$\langle\mathbf{l}\cdot\mathbf{s}\rangle$ is the expectation value of
the spin--orbit coupling operator.

The effective spin projection relevant to the XMCD spin sum rule is
defined as

\begin{equation}
    \langle S_{\rm eff}\rangle
    =
    \langle S_z\rangle
    +
    \frac{7}{2}\langle T_z\rangle .
    \label{eq:Seff}
\end{equation}

Table~\ref{tabS1} shows the state-resolved Crispy output for
$D_\sigma=D_\tau=0$, corresponding to the $O_h$ crystal-field limit,
as a representative example of the quantities entering the thermal
averages. Crispy denotes projections along the quantization axis by
$S_k$, $L_k$, $J_k$, and $T_k$, corresponding here to
$S_z$, $L_z$, $J_z$, and $T_z$, respectively.

\begin{table*}[htbp]
\caption{
State-resolved output from Crispy of the Co$^{2+}$ crystal-field multiplet
calculation for the $O_h$ crystal-field limit
($D_\sigma=D_\tau=0$) at $T=300$~K and $\mu_0H=6$~T.
The calculation used $10Dq=0.9$~eV, $F_k=0.75$, $G_k=0.78$, and
$\zeta=1$. $dZ_i$ denotes the normalized Boltzmann weight of each
state.
}
\label{tabS1}
\centering
\scriptsize
\setlength{\tabcolsep}{4.0pt}
\renewcommand{\arraystretch}{1.2}

\begin{tabular}{c
                c
                ccc
                cccc
                c
                cc
                c}
\hline\hline

State
& $E$
& $\langle S^2\rangle$
& $\langle L^2\rangle$
& $\langle J^2\rangle$
& $\langle S_k\rangle$
& $\langle L_k\rangle$
& $\langle J_k\rangle$
& $\langle T_k\rangle$
& $\langle\mathbf{l}\cdot\mathbf{s}\rangle$
& $\langle N_{2p}\rangle$
& $\langle N_{3d}\rangle$
& $dZ_i$ \\

& (eV) & & & & & & & & & & & \\

\hline

1
& $-2.88446$
& 3.7452 & 11.8064 & 23.0734
& $-0.8601$ & $-0.6303$ & $-1.4904$ & 0.0067
& $-1.3726$ & 6 & 7 & $3.68\times10^{-1}$ \\

2
& $-2.88290$
& 3.7452 & 11.8062 & 23.0682
& 0.7981 & 0.5310 & 1.3290 & $-0.0056$
& $-1.3717$ & 6 & 7 & $3.47\times10^{-1}$ \\

3
& $-2.84057$
& 3.7436 & 11.7937 & 19.4356
& $-1.1069$ & 0.2555 & $-0.8514$ & 0.0175
& $-0.7748$ & 6 & 7 & $6.75\times10^{-2}$ \\

4
& $-2.84001$
& 3.7437 & 11.7936 & 19.4270
& $-0.2884$ & 0.3066 & 0.0183 & $-0.0433$
& $-0.7720$ & 6 & 7 & $6.60\times10^{-2}$ \\

5
& $-2.83977$
& 3.7436 & 11.7938 & 19.4318
& 0.3274 & $-0.2461$ & 0.0813 & 0.0401
& $-0.7746$ & 6 & 7 & $6.54\times10^{-2}$ \\

6
& $-2.83924$
& 3.7437 & 11.7937 & 19.4206
& 1.0894 & $-0.2923$ & 0.7970 & $-0.0159$
& $-0.7714$ & 6 & 7 & $6.41\times10^{-2}$ \\

7
& $-2.76903$
& 3.7351 & 11.7734 & 13.1147
& $-0.8362$ & 0.5377 & $-0.2985$ & 0.1229
& 0.1857 & 6 & 7 & $4.24\times10^{-3}$ \\

8
& $-2.76886$
& 3.7352 & 11.7733 & 13.1056
& $-0.6830$ & 0.6888 & 0.0058 & 0.0185
& 0.1882 & 6 & 7 & $4.21\times10^{-3}$ \\

9
& $-2.76842$
& 3.7351 & 11.7735 & 13.1031
& 0.5949 & $-0.5946$ & 0.0003 & $-0.0202$
& 0.1868 & 6 & 7 & $4.14\times10^{-3}$ \\

10
& $-2.76822$
& 3.7351 & 11.7734 & 13.0963
& 0.8537 & $-0.5117$ & 0.3420 & $-0.1247$
& 0.1887 & 6 & 7 & $4.11\times10^{-3}$ \\

11
& $-2.75487$
& 3.7483 & 11.7543 & 11.5038
& $-0.4410$ & 0.5472 & 0.1062 & $-0.1199$
& 0.6095 & 6 & 7 & $2.45\times10^{-3}$ \\

12
& $-2.75459$
& 3.7483 & 11.7544 & 11.5113
& 0.5521 & $-0.6178$ & $-0.0658$ & 0.1234
& 0.6083 & 6 & 7 & $2.42\times10^{-3}$ \\

\hline\hline
\end{tabular}
\end{table*}

\FloatBarrier

The same thermal averaging procedure was applied to the calculated
output for each value of $D_\sigma$, with the resulting quantities
summarized in Table~\ref{tabS2}.

\begin{table*}[htbp]
\caption{
Thermally averaged quantities obtained from the Co$^{2+}$ atomic
multiplet calculations using Crispy at $T=300$~K and $\mu_0H=6$~T
for different values of the trigonal crystal-field parameter
$D_\sigma$. In all calculations, $D_\tau=0$, $10Dq=0.9$~eV,
$F_k=0.75$, $G_k=0.78$, and $\zeta=1$ were kept fixed.
The case $D_\sigma=D_\tau=0$ corresponds to the $O_h$ limit.
}
\label{tabS2}
\centering
\scriptsize
\setlength{\tabcolsep}{4.2pt}
\renewcommand{\arraystretch}{1.2}

\begin{tabular}{c|ccccccccccc}
\hline\hline

$D_\sigma$
& $\langle L^2\rangle$
& $\langle S^2\rangle$
& $\langle J^2\rangle$
& $\langle L_z\rangle$
& $\langle S_z\rangle$
& $\langle J_z\rangle$
& $\langle T_z\rangle$
& $\langle S_{\rm eff}\rangle$
& $\langle \mathbf{l}\cdot\mathbf{s}\rangle$
& $\dfrac{\langle L_z\rangle}{\langle S_z\rangle}$
& $\dfrac{\langle L_z\rangle}{\langle S_{\rm eff}\rangle}$ \\

(eV) & & & & & & & & & & & \\

\hline

$-0.15$
& 11.3184 & 3.7450 & 17.9877
& $-0.0144$ & $-0.0350$ & $-0.0494$ & 0.00352
& $-0.0227$ & $-0.6200$ & 0.411 & 0.633 \\

$-0.10$
& 11.5485 & 3.7436 & 19.3782
& $-0.0171$ & $-0.0323$ & $-0.0494$ & 0.00257
& $-0.0233$ & $-0.8094$ & 0.528 & 0.731 \\

$-0.05$
& 11.7366 & 3.7446 & 21.1274
& $-0.0281$ & $-0.0335$ & $-0.0616$ & 0.00155
& $-0.0281$ & $-1.0647$ & 0.837 & 0.998 \\

$0$
& 11.7971 & 3.7430 & 21.8804
& $-0.0446$ & $-0.0423$ & $-0.0869$ & 0.00046
& $-0.0407$ & $-1.1784$ & 1.055 & 1.096 \\

$+0.05$
& 11.7546 & 3.7418 & 21.4463
& $-0.0524$ & $-0.0486$ & $-0.1010$ & $-0.00049$
& $-0.0503$ & $-1.1143$ & 1.079 & 1.042 \\

$+0.10$
& 11.6714 & 3.7414 & 20.7222
& $-0.0516$ & $-0.0503$ & $-0.1018$ & $-0.00049$
& $-0.0520$ & $-1.0096$ & 1.025 & 0.991 \\

$+0.15$
& 11.5634 & 3.7421 & 19.9922
& $-0.0457$ & $-0.0495$ & $-0.0952$ & $-0.00178$
& $-0.0557$ & $-0.9087$ & 0.923 & 0.820 \\

\hline\hline
\end{tabular}
\end{table*}

\FloatBarrier

A notable result is that $\langle S^2\rangle$ remains nearly constant
at 3.74 throughout the investigated $D_\sigma$ range, close to
$S(S+1)=3.75$ for $S=3/2$, confirming that the high-spin Co$^{2+}$
state is preserved. In contrast, $\langle L^2\rangle$,
$\langle J^2\rangle$, and
$\langle\mathbf{l}\cdot\mathbf{s}\rangle$ vary appreciably with
$D_\sigma$, reflecting a redistribution of the orbital and spin--orbit
character by the trigonal crystal field. In the $O_h$ limit,
$\langle L_z\rangle/\langle S_{\rm eff}\rangle=1.096$, compared with
0.998 and 1.042 for $D_\sigma=-0.05$ and $+0.05$~eV, respectively.
Thus, throughout the weakly trigonal regime giving the best spectral
agreement, the orbital and effective-spin projections remain of
comparable magnitude. The corresponding $\langle T_z\rangle$ values
are small, indicating only a modest magnetic-dipole contribution.

For comparison, the experimental XMCD sum-rule analysis gives
\begin{equation}
    \langle L_z\rangle^{\rm exp}=0.034(7),
    \qquad
    \langle S_{\rm eff}\rangle^{\rm exp}=0.024(5),
    \qquad
    \frac{\langle L_z\rangle^{\rm exp}}
         {\langle S_{\rm eff}\rangle^{\rm exp}}=1.38(15).
\end{equation}
The quoted uncertainties represent conservative estimates of the
systematic sensitivity to reasonable variations of the spectral
processing parameters. The absolute values are primarily sensitive to
the linear pre-edge background, whereas the uncertainty of their ratio
is dominated by the endpoint of the $L_3$ dichroic integral $p$ in the
inter-edge region. The nominal analysis uses $p$ at 788~eV and the
total dichroic integral $q$ at 800~eV. Although the experimental ratio
is somewhat larger than the calculated values, both consistently
indicate a substantial unquenched orbital contribution.

To compare the expectation values with magnetic moments, we use
$m_L=\mu_{\rm B}|\langle L_z\rangle|$ and
$m_{S_{\rm eff}}=2\mu_{\rm B}|\langle S_{\rm eff}\rangle|$.
Consequently,
$\langle L_z\rangle/\langle S_{\rm eff}\rangle=1.38(15)$ corresponds
to $m_L/m_{S_{\rm eff}}=0.69(8)$ before any additional correction.
For Co$^{2+}$, the effective-spin sum rule requires a correction for
the incomplete separation of the $L_3$ and $L_2$ edges and associated
core-hole multiplet effects. Following Piamonteze \textit{et al.}
\cite{Piamonteze2009}, the calculated spin-sum-rule value for high-spin
Co$^{2+}$ ($3d^7$) with atomic spin--orbit coupling and
$10Dq=0.9$~eV is approximately 0.925 of the exact effective-spin
expectation value, corresponding to
$C_S\simeq1/0.925\simeq1.08$. A similar correction has previously
been applied to high-spin Co$^{2+}$ compounds \cite{Lee_2015}.
The resulting experimental moments are therefore
\begin{equation}
    m_L^{\rm exp}=0.034(7)\,\mu_{\rm B},
    \qquad
    m_{S_{\rm eff}}^{\rm exp,corr}=0.052(11)\,\mu_{\rm B},
    \qquad
    \frac{m_L^{\rm exp}}
         {m_{S_{\rm eff}}^{\rm exp,corr}}=0.64(7).
\end{equation}
The orbital-to-effective-spin magnetic-moment ratio is therefore large
but physically reasonable for high-spin Co$^{2+}$.

The absolute moment obtained from the single-ion calculation is somewhat
larger than the experimental value because the Hamiltonian does not include
the strong antiferromagnetic exchange interactions in
NaCdCo$_2$F$_7$. In the $O_h$ limit at 300~K and 6~T, the calculation
gives $m_L=0.0446~\mu_{\rm B}$ and
$m_{S_{\rm eff}}=0.0814~\mu_{\rm B}$, corresponding to a total moment
$m_{\rm calc}=0.126~ \mu_{\rm B}$. As a simple estimate of the effect of
antiferromagnetic correlations, the experimentally determined
$\theta_{\rm CW}=-108(1)$~K gives a Curie--Weiss reduction of the
susceptibility relative to the noninteracting Curie response by
\begin{equation}
\frac{\chi_{\rm CW}}{\chi_{\rm C}}
=
\frac{T}{T-\theta_{\rm CW}}
=
\frac{300}{300+108}
\simeq0.735 .
\end{equation}
Applying this factor to the calculated total moment gives
$m_{\rm calc}^{\rm CW} \simeq 0.735m_{\rm calc}\simeq0.0926,\mu_{\rm B}$, in close agreement
with the experimental magnetization measured at the same temperature
and field,
$m_{\rm VSM}^{300 {\rm K} ,6{\rm T}}=0.092 \mu_{\rm B}$.
This comparison is intended only as a simple mean-field estimate, since
exchange interactions are not included explicitly in the single-ion
calculation.

Taken together, the spectral comparison and calculated expectation
values support a predominantly octahedral crystal-field environment
with at most a weak trigonal perturbation whose magnitude and sign
cannot be uniquely determined from the present data. Both the
calculations and XMCD sum-rule analysis consistently establish a
substantial unquenched orbital contribution to the Co$^{2+}$ magnetic
moment, supporting the use of the $O_h$ calculation as the minimal
description in the main text.

\bibliography{bibliography}